\documentclass[twocolumn,nopacs,floatfix,amsmath,nofootinbib,amssymb,preprintnumbers]{revtex4-2}
\usepackage{graphicx,color,dcolumn,booktabs,bm}
\usepackage{txfonts}
\usepackage{xcolor}
\usepackage{slashed}
\usepackage[colorlinks,
            citecolor=blue,
            anchorcolor=red,
            menucolor=red,
            linkcolor=red,
            filecolor=red,
            runcolor=red,
            urlcolor=blue,
            frenchlinks=red]{hyperref}

\begin{document}

\preprint{
	\vbox{
		\hbox{ADP-24-11/T1250}
}}

\title{Pion photoproduction of nucleon excited states with Hamiltonian effective field theory}

\author{Yu Zhuge$^{1,2,3}$}
\author{Zhan-Wei Liu$^{1,2,3}$}\email{liuzhanwei@lzu.edu.cn}
\author{Derek~B.~Leinweber$^4$}\email{derek.leinweber@adelaide.edu.au}
\author{Anthony W. Thomas$^4$}\email{anthony.thomas@adelaide.edu.au}
\affiliation{
$^1$School of Physical Science and Technology, Lanzhou University, Lanzhou 730000, China\\
$^2$Research Center for Hadron and CSR Physics, Lanzhou University and Institute of Modern Physics of CAS, Lanzhou 730000, China\\
$^3$Lanzhou Center for Theoretical Physics, Key Laboratory of Theoretical Physics of Gansu Province, Key Laboratory of Quantum Theory and Applications of MoE, and MoE Frontiers Science Center for Rare Isotopes, Lanzhou University, Lanzhou 730000, China\\
$^4$CSSM,
%ARC Special Research Centre for the Subatomic Structure of
%  Matter,
  Department of Physics, University of Adelaide, South
  Australia 5005, Australia
}

\begin{abstract}
We refine our previous calculation of multipole amplitude $E_{0+}$ for pion photoproduction process, $\gamma N\rightarrow\pi N$. The treatment of final-state interactions is based upon an earlier analysis of pion-nucleon scattering within Hamiltonian effective field theory, supplemented by incorporating contributions from the $N^*(1650)$ and the $K\Lambda$ coupled channel. The contribution from the bare state corresponding to the $N^*(1650)$ significantly enhances our results. Additionally, we also compute the multipole amplitude $M_{1-}$, which is of direct relevance to the Roper resonance. The results are comparable with other dynamical coupled channel models, even though the contribution from the bare state (interpreted as a 2$s$ excitation) in this channel is small because of its large mass.
\end{abstract}

\maketitle

\section{introduction}\label{sec1}
An important challenge in hadron physics is to understand quantum chromodynamics (QCD) in the non-perturbative regime. The numerous four star $N$, $\Delta$ and $\Lambda$ resonances~\cite{Workman:2022ynf} constitute an ideal arena within which to study these non-perturbative effects.

The earliest reaction used to study these baryon resonances and extract information on their properties was $\pi N$ scattering~\cite{Klempt:2009pi, Crede_2013, Thiel:2022xtb}. By partial-wave analysis, masses, widths and decay branching ratios of these resonances can be determined~\cite{Cutkosky:1979fy, Arndt:1995bj, morsch2001, Arndt:2002xv, Arndt:2006bf}. The quark model is the simplest and most effective phenomenological model for studying the baryon spectrum. Using this model many baryon resonances were successfully predicted~\cite{Capstick:1986ter, Loring:2001kv, Loring:2001kx, Loring:2001ky, Melde:2008yr, Ferretti:2011zz}. On the other hand, some of the predicted states have not been discovered in experiments~\cite{Lyu:2024qgc, Wang:2024jyk}. 

Koniuk and Isgur~\cite{Koniuk:1979vy} suggested that these ``missing'' states couple weakly to the $\pi N$ channel, making them difficult to detect in $\pi N$ scattering experiments. Therefore, photoproduction and electroproduction are essential mechanism for discovering new baryon resonances. Additionally, these reactions provide more information about the baryon resonances, such as their helicity amplitudes and form factors, which helps us better understand their internal structure~\cite{Ramalho:2023hqd}. Reviews~\cite{Klempt:2009pi, Crede_2013, Thiel:2022xtb, Ramalho:2023hqd} provide more detailed information about the experimental progress in light baryon spectroscopy.

Based on chiral perturbation theory (ChPT), photoproduction in the low energy region is understood well~\cite{Bernard:1991rt, Bernard:1992nc, Bernard:1994gm, Bernard:2007zu, Ruic:2011wf, Mai:2012wy, Hilt:2013uf}. At higher energy, pion- and photon-induced reactions have been extensively analyzed through partial wave analysis in dynamical coupled-channel models, like the ANL-Osaka model or EBAC~\cite{Julia-Diaz:2006ios, Matsuyama:2006rp, Juli_D_az_2008, Kamano:2009im, Suzuki:2010yn, Sandorfi:2010uv, Kamano:2013iva, Kamano:2016bgm, kamano2019anlosaka}, the J{\"u}lich-Bonn model~\cite{Doring:2009yv, Doring:2009bi, Ronchen:2014cna, Ronchen:2014ooa, Ronchen:2015vfa, Ronchen:2016sjn, Anisovich:2016vzt, Ronchen:2018ury, Ronchen:2022hqk} and the Bonn-Gatchina model~\cite{Anisovich:2008zz, Anisovich:2012ct, Anisovich:2011fc, Sarantsev:2016fac, Anisovich:2016vzt}. These models consider a large number of meson-baryon coupled channels, enabling a comprehensive analysis of scattering data for light baryons. Resonance information is extracted from this analysis. By defining the interaction vertex functions between resonances and coupled channels, the helicity amplitudes and form factors of the resonances can also be obtained.

Lattice QCD is the only first-principles theory for studying light baryons and their excited states. It enables us to calculate the masses and other properties of hadrons on a finite-sized lattice, using a range of quark masses. Ideally, the lattice spacing should be infinitesimally small with a continuum limit, but this is constrained by current computational capabilities. Therefore, methods are required to extrapolate the lattice QCD results to the continuum limit and the physical mass region.

L{\"u}scher's method is an effective tool for linking the finite volume spectrum in lattice QCD at physical quark masses with infinite volume experimental observables~\cite{Luscher:1985dn, Luscher:1986pf, Luscher:1990ux}. The Hamiltonian effective field theory (HEFT) is also a useful approach for connecting lattice QCD results with experimental data. By constructing the Hamiltonian of the system and solving the coupled channel equations, one can obtain scattering observables such as inelasticities, phase shifts, and cross sections. By applying the Hamiltonian in finite volume, the energy spectrum can be obtained and compared with the lattice QCD spectrum, helping us to understand the structure of baryon resonances. HEFT has been used to achieve a number of insights into the structure of nucleon, $\Delta$, and $\Lambda$ resonances~\cite{Hall:2013qba, Wu:2014vma, Hall:2014uca, Liu:2015ktc, Liu:2016wxq, Liu:2016uzk, Wu:2017qve, Li:2019qvh, Liu:2020foc, Li_2021, Abell:2021awi, Abell:2023nex, Liu:2023xvy}. 

This approach has enhanced our understanding of the internal structure of these resonances, including the fact that a number of them are not primarily three-quark states but dynamically generated. The parameters describing the coupling of the resonances to coupled meson-baryon channels are obtained by fitting to scattering data and validated against the lattice QCD mass spectrum. Using these parameters, we can further analyze the electromagnetic properties of the baryon excited states in reactions such as pion photoproduction, with minimal model-dependent assumptions.

Our previous study~\cite{Liu:2015ktc} on the $N^*(1535)$  used a single bare state in the Hamiltonian and fitted the $\pi N$ scattering in the $S_{11}$ partial wave to obtain the model parameters. Recently two bare basis states of low-lying odd-parity nucleon resonances have been considered in HEFT~\cite{Abell:2023nex}, which effectively described both the $S_{11}$ $\pi N$ scattering and the finite-volume spectrum observed in lattice QCD. Considering the interference between the $N^*(1535)$ and $N^*(1650)$ resonances, as well as their coupling to the newly added $K\Lambda$ channel, we can refine our calculation of multipole amplitude $E_{0+}$. 

Since the discovery of the Roper resonance, it has proven a major challenge to theoretical interpretation. For example, there is the issue of its mass inversion: it has a lower mass than the $N^*(1535)$, which contradicts the predictions of the oscillator-inspired quark model. In a previous study of the 
$N^*(1440)$~\cite{Wu:2017qve}, we included three coupled channels, $\pi N$, $\pi\Delta$ and $\sigma N$. The $N^*(1440)$ was found to be generated by strong rescattering between the coupled meson-baryon channels, $\pi N$, $\sigma N$ and $\pi\Delta$, with only a small quark-model-like state component, having a mass around 2 GeV. By fitting the scattering T-matrix of these hadronic channels, we obtained the coupling constants and cutoff parameters of the interaction vertices. The parameter sets are described in Table I in Ref.~\cite{Wu:2017qve}. Two scenarios were considered there to describe the structure of the $N^*(1440)$, but only Scenario I was consistent with the lattice QCD results. Thus, in this work, we use the parameters from Scenario I to calculate the pion photoproduction in the $P_{11}$ partial wave.

This paper is organized as follows. In Sec.~\ref{sec2}, we outline the framework of calculating the multipole amplitudes by combining the  electromagnetic transitions with the final-state interactions (FSI) effects. The numerical results and some discussion are provided in Sec.~\ref{ND}. A brief summary is presented in Sec.~\ref{summary}.

%%%%%%%%%%%%%%%%%%%%%%%%%%%%%%%%%%%%%%%%%%%
\section{framework}\label{sec2}
The calculation of the scattering T-matrix for the reaction $\gamma N\rightarrow\pi N$ may be split into two parts, the electromagnetic transitions $\gamma N\overset{\rm EM}{\rightarrow}\alpha$ and the FSI $\alpha\overset{\rm FSI}{\rightarrow}\pi N$, where $\alpha$ is the coupled channel involved. The electromagnetic amplitudes are derived from well-established effective Lagrangians. For the FSI we employ the half-off-shell scattering amplitudes generated in our earlier HEFT studies.

%%%%%%%%%%%%%%%
\subsection{Electromagnetic transitions without FSI}
\label{emt}
\begin{figure*}[htbp]
		\centering
		\includegraphics[width=\textwidth]{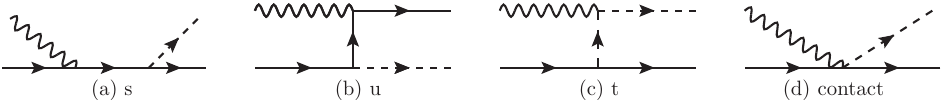}
		\caption{Tree-level diagrams for the pure electromagnetic amplitude of the $\gamma N\to \alpha$ process without FSI: (a) s channel, (b) u channel, (c) t channel, (d) the contact term. The solid, wiggly, and dashed lines represent the baryons, photons and mesons,  respectively.}\label{sut}
\end{figure*}

The effective Lagrangians shown in Appendix~\ref{app1} used in $\gamma N\overset{\rm EM}{\rightarrow}\alpha$ process are taken from  the ANL-Osaka model~\cite{Matsuyama:2006rp, Kamano:2013iva}. The complete scattering amplitude of $\gamma N\rightarrow\alpha$ with explicit spin directions has the following from:
\begin{eqnarray}
\label{fullem}
		&&\mathcal M_{\alpha,\gamma N}(s_z^{\prime {\rm B}_\alpha},\lambda_N, \lambda_\gamma;\vec{k},\vec{q})\notag\\
		&\equiv&
		\langle {\rm M}_\alpha(\vec{k}),\, {\rm B}_\alpha(\vec{p}\,',s_z^{\prime {\rm B}_\alpha})|\,H^{\rm EM}\,|\gamma(\vec{q},\lambda_\gamma),\,N(\vec{p},-\lambda_N)\rangle
		\notag\\
		&=&
		\frac{1}{(2\pi)^3} \frac{1}{\sqrt{2\omega_{{\rm M}_\alpha}(k)}}\frac{1}{\sqrt{2|\vec q|}}\,\,\bar{u}_{{\rm B}_\alpha}(s_z^{\prime {\rm B}_\alpha})\, \hat{\mathcal{M}}_{\alpha,\gamma N}\, u_N(-\lambda_N) \, ,
\end{eqnarray}
where ${\rm M}_\alpha$ and ${\rm B}_\alpha$ refer to the meson and the baryon in channel $\alpha$, respectively. $\hat{\mathcal{M}}_{\alpha,\gamma N}$ is defined as follows
\begin{equation}
\label{pnM}
\hat{\mathcal{M}}_{\alpha,\gamma N}\equiv
\left(\begin{array}{cc}
\hat{\mathcal{M}}_{\alpha, \gamma p}&0\\
0&\hat{\mathcal{M}}_{\alpha, \gamma n}
\end{array}\right),
\end{equation}
where the amplitudes 
$\hat{\mathcal{M}}_{\alpha, \gamma p}$ and $\hat{\mathcal{M}}_{\alpha, \gamma n}$ representing proton and neutron targets and  listed in Appendix~\ref{app1}, incorporate contributions from the bare resonance states through the $s$- and $u$-channels as shown in Fig.~\ref{sut}. The spin of the $\Delta$ is $\frac{3}{2}$, and its spin wave function is obtained by coupling the wave functions with spin 1 and spin $\frac{1}{2}$~\cite{Hemmert:1997ye, Nozawa1990ADM}
\begin{align}
	u_{\mu,\Delta}(s^{\prime\Delta}_z)=\sum_{\lambda,s}(1\lambda\frac{1}{2}s|\frac{3}{2}s^{\prime\Delta}_z)e_\mu(\lambda)u(s),
\end{align}
where $e_\mu(\pm)=\mp\frac{1}{\sqrt{2}}(0,1,\pm i,0)$ and $e_\mu(0)=(0,0,0,1)$ represent the wave functions for spin 1, and $u(\frac{1}{2})=(1,0,0,0)$ and $u(-\frac{1}{2})=(0,1,0,0)$ represent the wave functions for spin $\frac{1}{2}$. With these, we can write the expressions for the four spin components of the $\Delta$ particle (only two are given here)
\begin{align}
\label{deltawavefuc}
	u_{\mu,\Delta}(\frac{3}{2})&=(11\frac{1}{2}\frac{1}{2}|\frac{3}{2}\frac{3}{2})e_\mu(1)u(\frac{1}{2})=e_\mu(1)u(\frac{1}{2}),\notag\\
	u_{\mu,\Delta}(\frac{1}{2})&=(11\frac{1}{2}-\frac{1}{2}|\frac{3}{2}\frac{1}{2})e_\mu(1)u(-\frac{1}{2})+(10\frac{1}{2}\frac{1}{2}|\frac{3}{2}\frac{1}{2})e_\mu(0)u(\frac{1}{2})\notag\\
	&=\sqrt{\frac{1}{3}}e_\mu(1)u(-\frac{1}{2})+\sqrt{\frac{2}{3}}e_\mu(0)u(\frac{1}{2}) \, .
\end{align}
%
%where $e_\mu(\lambda)$ contracts with the four-momentum in the Feynman amplitude, while $u(s)$ contracts with the Dirac matrix. 
We employ the helicity-$LSJ$ mixed-representation, as defined in Ref.~\cite{Matsuyama:2006rp}, to derive the partial wave amplitude. In our case, it is as follows~\cite{Jacob:1959at, Richman:153636}
\begin{align}
\label{JLS}
	V^{JLS;\lambda_\gamma\lambda_N}_{\alpha,\gamma N}(k,q)={}&4\pi\sqrt{\frac{2J+1}{4\pi}}\sum_{\begin{smallmatrix}mm_s\\m_1m_2\end{smallmatrix}}(lmsm_s|JM)(s_1m_1s_2m_2|sm_s)\notag\\
	&\times\int d\Omega\,Y^*_{lm}(\theta,\phi)\mathcal M_{\alpha,\gamma N}(s_z^{\prime {\rm B}_\alpha},\lambda_N, \lambda_\gamma;\vec{k},\vec{q}).
\end{align}
Here we have chosen the direction of the photon momentum, $\vec{q}$, to lie along the $z$ axis. $\theta$ is the angle between $\vec k$ and $\vec q$. $s_1$ and $s_2$ represent the spin of the meson and the baryon in channel $\alpha$, respectively. We have $\lambda_\gamma=1$ and $\lambda_N=\frac{1}{2}$ for the $S_{11}$ and $P_{11}$ partial waves. The full amplitude $\mathcal M_{\alpha,\gamma N}(s_z^{\prime {\rm B}_\alpha},\lambda_N, \lambda_\gamma;\vec{k},\vec{q})$ is defined as Eq.~(\ref{fullem}).

The $N^*(1535)$ and $N^*(1650)$ both have negative parity. We consider three coupled channels, $\pi N$, $\eta N$ and $K \Lambda$. The final potential,  $V^{JLS;\lambda_\gamma\lambda_N}_{\alpha,\gamma N}(k,q)$, in this case is as follows
\begin{align}
	&V^{JLS;\lambda_\gamma,\lambda_N}_{\alpha,\gamma N}(k,q)\notag\\
	={}&4\pi\sqrt{\frac{2J+1}{4\pi}}\int_{0}^{2\pi}d\varphi\int_{0}^{\pi}d\theta\,\sin\theta\,Y^*_{0,0}(\theta,\phi)\notag\\
	&\times\mathcal{M}_{\alpha,\gamma N}(s^{\prime N}_z=\frac{1}{2},\lambda_N,\lambda_\gamma;\vec{k},\vec{q}) \, .
\end{align}

The $N^*(1440)$ has positive parity. The final potentials for the three coupled channels, $\pi N$, $\sigma N$ and $\pi\Delta$, are:
\begin{align}
	&V^{JLS;\lambda_\gamma,\lambda_N}_{\pi N,\gamma N}(k,q)\notag\\
	={}&4\pi\sqrt{\frac{2J+1}{4\pi}}  (10\frac{1}{2}\frac{1}{2}|\frac{1}{2}\frac{1}{2})\int_{0}^{2\pi}d\varphi\int_{0}^{\pi}d\theta\,\sin\theta\,Y^*_{1,0}(\theta,\phi)\notag\\
	&\times\mathcal{M}_{\pi N,\gamma N}(s^{\prime N}_z=\frac{1}{2},\lambda_N,\lambda_\gamma;\vec{k},\vec{q}),
\end{align}
\begin{align}
	&V^{JLS;\lambda_\gamma,\lambda_N}_{\sigma N,\gamma N}(k,q)\notag\\
	={}&4\pi\sqrt{\frac{2J+1}{4\pi}}\int_{0}^{2\pi}d\varphi\int_{0}^{\pi}d\theta\,\sin\theta\,Y^*_{0,0}(\theta,\phi)\notag\\
	&\times\mathcal{M}_{\sigma N,\gamma N}(s^{\prime N}_z=\frac{1}{2},\lambda_N,\lambda_\gamma;\vec{k},\vec{q}),
\end{align}
and
\begin{align}
	&V^{JLS;\lambda_\gamma,\lambda_N}_{\pi\Delta,\gamma N}(k,q)\notag\\
	={}&4\pi\sqrt{\frac{2J+1}{4\pi}}(10\frac{3}{2}\frac{1}{2}|\frac{1}{2}\frac{1}{2})\int_{0}^{2\pi}d\varphi\int_{0}^{\pi}d\theta\,\sin\theta Y^*_{1,0}(\theta,\phi)\notag\\
	&\times\mathcal{M}_{\pi\Delta,\gamma N}(s^{\prime \Delta}_z=\frac{1}{2},\lambda_N,\lambda_\gamma;\vec{k},\vec{q}) \, .
\end{align}

When calculating the partial wave amplitude, we introduce a form factor into the electromagnetic potential. The form factors are taken to be $u^{\rm em}_{-}(k)=\left(1+\frac{k^2}{\Lambda_{-}^{{\rm em}\,2}}\right)^{-2}$ and $u^{\rm em}_+(k)=\exp(-k^2/\Lambda_{+}^{{\rm em}\,2})$ for the $S_{11}$ and $P_{11}$ partial waves, respectively, similar to those used in the FSI treatment~\cite{Wu:2017qve, Abell:2023nex}. In the electromagnetic transition process, we include contributions from the bare excited states of the nucleon. These act as intermediate propagators in the $s$- and $u$- channels. The Lagrangians involving the resonances and the corresponding amplitudes are listed in Appendix~\ref{app1}.

%%%%%%%%%%%%%%%%%%%%%%%%%%%%%%%%%%%%%%%%%%%%%%%%%%%%%%%%%%%%%%%%%%%%%%%%%%%%%%%%%%%%%%%%%%%%
\subsection{FSI within HEFT}
The interacting Hamiltonian developed in our HEFT analysis of these reonances contains two parts, $H_I=g+v$. The first term, $g$, represents the vertex interaction between the bare state $N^*$ and coupled channel $\alpha$
\begin{align}
    g=\sum_{\alpha, i}\int d^3\vec{k}\left\{|\alpha(\vec k)\rangle\, G_\alpha^\dagger(k)\,\langle N^*_i| + |N_i\rangle\, G_\alpha(k)\,\langle\alpha(\vec k)| \right\} \, .
\end{align}
The second term, $v$, represents the interaction between two-particles basis states
\begin{align}
    v=\sum_{\alpha,\beta}\int d^3\vec{k}d^3\vec{k^\prime}|\alpha(\vec k)\rangle\, V_{\alpha,\beta}(k,k')\, \langle\beta(\vec{k}')| \, .
\end{align}
The interaction function $G_\alpha(k)$ for the odd parity resonances $N^*(1535)$ and $N^*(1650)$ takes the form
\begin{equation}
  G_{\alpha}^{N^{*-}_i}(k)=\frac{\sqrt{3} g_{\alpha}^{N^{*-}_i}}{2\pi f_{\pi}} \sqrt{\omega_{\text{M}_{\alpha}}(k)} u^-(k) \, ,
\end{equation}
where $\text{M}_{\alpha}$ refers to the meson in channel $\alpha$, giving $\omega_{\text{M}_{\alpha}}(k) = \sqrt{k^2 + m_{\text{M}_{\alpha}}^2}$, $f_{\pi} = 92.4$ MeV. $u^-(k)$ is the dipole regulator of the form $u^-(k)=\left(1+\frac{k^2}{\Lambda^2}\right)^{-2}$~\cite{Abell:2023nex}. For the even parity resonance $N^*(1440)$, $G_{\alpha}^{N^{*+}_i}(k)$ takes the form
\begin{equation}
  G_{\alpha}^{N^{*+}_i}(k)=\frac{g_\alpha^{N^{*+}_i}}{2\pi}\left(\frac{k}{f_\pi}\right)^{l_\alpha}\frac{u^+(k)}{\sqrt{\omega_{\rm M_\alpha}(k)}} \, ,
\end{equation}
where $l_\alpha$ is the orbital angular momentum in channel $\alpha$.
%$l$ is $1$ for $\pi N$ and $\pi\Delta$, while it is $0$ for $\sigma N$. 
$u^+(k)$ takes the exponential form, $u^+(k)=\exp\left(-k^2/\Lambda^2\right)$~\cite{Wu:2017qve}. The two-to-two particle interaction,  $V_{\alpha,\beta}(k,k')$, for the odd parity resonances takes the form
\begin{equation}
  V^-_{\alpha,\beta}(k,k') = \frac{3v^-_{\alpha,\beta}}{4\pi^2 f_{\pi}^2}\tilde{u}(k)\tilde{u}(k') \, ,
\end{equation}
where $\tilde{u}(k)=u^-(k)[m_\pi+\omega_\pi(k)]/\omega_\pi(k)$. For the even parity resonance, it takes the form
\begin{equation}
    V^+_{\alpha,\beta}(k,k') = v^+_{\alpha,\beta}\frac{\bar{G}_\alpha(k)}{\sqrt{\omega_{\rm M_\alpha}(k)}}\frac{\bar{G}_\beta(k')}{\sqrt{\omega_{\rm M_\beta}(k')}} \, ,
\end{equation}
where $\bar{G}_\alpha(k) = G_{\alpha}^{N^+_i}(k)/g_\alpha^{N^+_i}$. Then one can obtain the T-matrix $T_{\alpha,\beta}$ by solving the relativistic, three-dimensional coupled channel equations
\begin{align}
  T_{\alpha,\beta}(k,k';E) &= \tilde{V}_{\alpha,\beta}(k,k',E) \nonumber\\
                          &+ \sum_{\gamma}\int dq\,q^2\frac{\tilde V_{\alpha,\gamma}(k,q,E) T_{\gamma,\beta}(q,k';E)}{E - \omega_\gamma(q) + i\epsilon} \, ,
\end{align}
where $\omega_{\gamma}(q)=\sqrt{q^2 + m_{\text{M}_{\gamma}}^2} + \sqrt{q^2 + m_{\text{B}_{\gamma}}^2}$. The coupled channel potential,  $\tilde{V}_{\alpha,\beta}$, is defined as 
\begin{equation}
  \tilde{V}_{\alpha,\beta}(k,k',E) = \sum_{i} \frac{G_{\alpha}^{N^*_i\,\dagger}(k)\,G_{\beta}^{N^*_i}(k')}{E - m^0_{N^*_i}} + V_{\alpha,\beta}(k,k') \, .\label{FSIb}
\end{equation}
%%%%%%%%%%%%%%%%%%%%%%%%%%%%%%%%%%%%%%%%%

%%%%%%%%%%%%%%%%%%%%%%%%%%%%%%%%%%%%%%
\subsection{The multipole amplitudes for $\gamma N\rightarrow\pi N$}
The scattering T-matrix for the reaction $\gamma N \rightarrow \pi N$ can be obtained by combining the electromagnetic potential,  $V^{JLS;\lambda_\gamma\lambda_N}_{\alpha,\gamma N}$, with the final-state interactions,  
$T_{\pi N,\alpha}$~\cite{Matsuyama:2006rp, Kamano:2013iva}.
\begin{align}
\label{finalT}
	T_{\pi N,\gamma N}^{\lambda_{\gamma},\lambda_{N}}(k,q;E)=&V_{\pi N,\gamma N}^{JLS;\lambda_{\gamma},\lambda_{N}}(k,q)+\sum_{\alpha}\int\mathrm{d}k'k^{\prime2}V_{\alpha,\gamma N}^{JLS;\lambda_{\gamma},\lambda_{N}}(k',q)\notag\\
	&\times\frac{1}{E-\omega_{\alpha}(k')+i\epsilon}T_{\pi N,\alpha}(k,k';E) \, .
\end{align}
The $S_{11}$ and $P_{11}$ partial wave amplitudes are related to the multipole amplitudes,  $E_{0+}$ and $M_{1-}$, respectively
\begin{equation}
    E_{0+}(E_{\rm cm})=\frac{\pi m_N\sqrt{\omega_{\pi}(k_{\rm on})\, q_{\rm on}}}{4E_{\rm cm}}
		T^{\lambda_\gamma,\lambda_N}_{\pi N,\gamma N}(k_{\rm on},q_{\rm on};E_{\rm cm}) \, ,
\end{equation}
\begin{equation}
    M_{1-}(E_{\rm cm})=\frac{\pi m_N\sqrt{\omega_\pi(k_{\rm on})q_{\rm on}}}{4E_{\rm cm}}T_{\pi N,\gamma N}^{\lambda_{\gamma},\lambda_{N}}(k_{\rm on},q_{\rm on};E_{\rm cm}) \, .
\end{equation}
%%%%%%%%%%%%%%%%%%%%%%%%%%%%%%%%%%%%%%%%%

%%%%%%%%%%%%%%%%%%%%%%%%%%%%%%%%%%%%%%%%%
\section{Numerical results and discussion}\label{ND}
The numerical results for the multipole amplitudes are calculated by combining the electromagnetic potential with the final-state interactions. The coupling constants associated with the resonances are determined by fitting to the single energy partial wave amplitudes provided by SAID~\cite{GWU}. We do not adjust the couplings appearing in the terms in the Lagrangian which do not involve the resonances. They are taken from  Refs.~\cite{Matsuyama:2006rp, Kamano:2013iva}, as listed in Table~\ref{couplingnotre}. As explained below, the exception are the constants  $f_{\eta NN}$ and $g_{\sigma NN}$. 
\begin{table}[tb]
	\caption{The couplings appearing in the Lagrangians not involving the resonances.}
	\centering
	\label{couplingnotre}
  \begin{ruledtabular}
	\begin{tabular}{crcr}

		Parameter &Value &Parameter &Value\\
		\hline
		$f_{\pi NN}$    &   1.003   &   $f_{KN\Sigma}$  &   0.969\\
		$g_{\gamma\rho\pi}$ &   0.128   &   $g^c_{\gamma K^*K}$    &   0.126\\
		$g_{\rho NN}$   &   4.724   &   $g^0_{\gamma K^*K}$    &   $-$0.192\\
		$g_{\gamma\rho\eta}$    &   1.150   &   $f_{\pi N\Delta}$    &  1.256\\
		$f_{KN\Lambda}$    &   $-$3.584   &   $f_{\pi \Delta\Delta}$    &  0.415\\
            $f_{\rho N\Delta}$    &    6.352    &    &\\

	\end{tabular}
  \end{ruledtabular}
\end{table}
%%%%%%%%%%%%%%%%%%%%%%%%%%%%%%%%%%%%%%%%%%%%%%%%%%%%%%%%%%%%%%%%%%%%%%%%%%%%%%%%

\subsection{The $S_{11}$ partial wave amplitude}
\begin{figure*}[htbp]
	\centering
	\begin{tabular}{cc}
		\includegraphics[height=6cm]{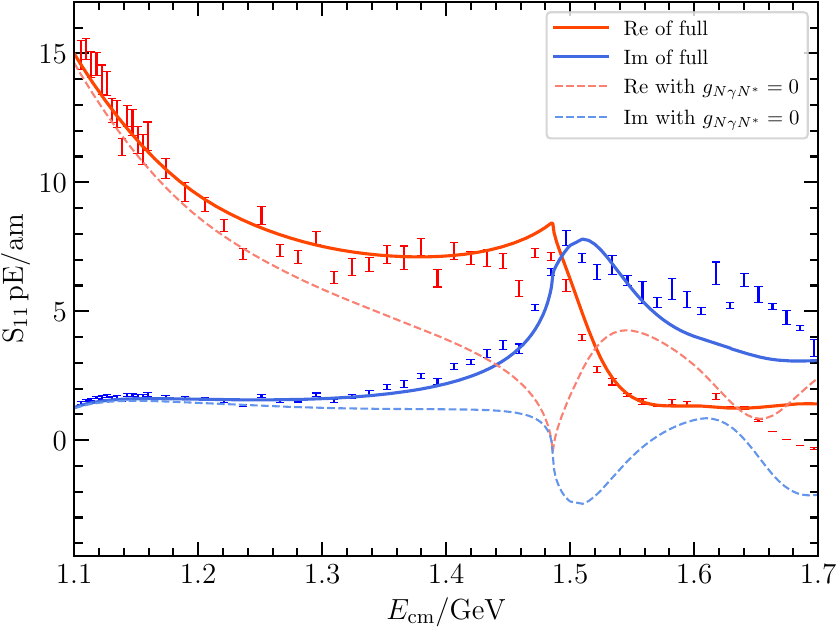}&
		\includegraphics[height=6cm]{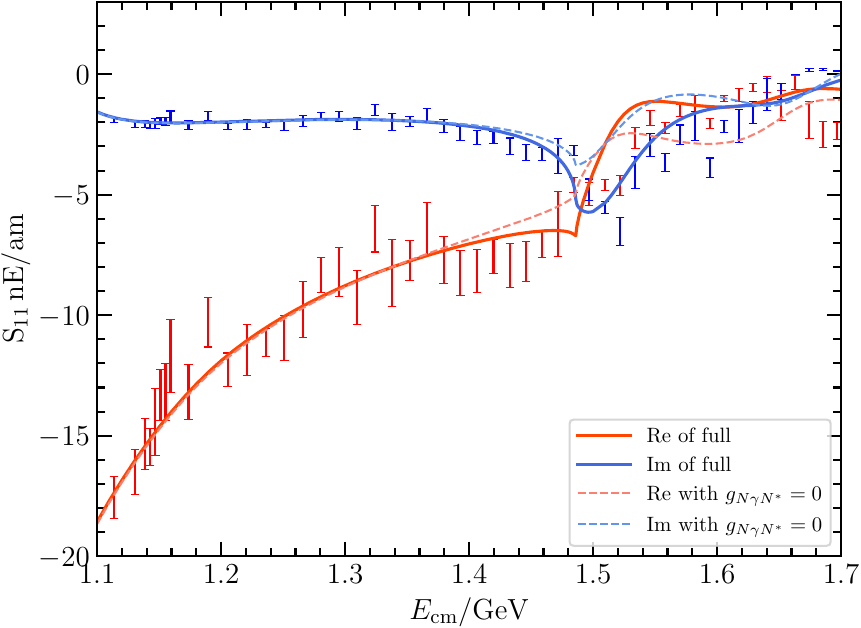}\\
		(a)&(b)
	\end{tabular}
	\caption{The fitted amplitude $E_{0+}$ in units of attometer. The solid and dashed lines, referring to the full and $\hat{g}_{N\gamma N^*}=0$ cases respectively, represent theoretical results combining the electromagnetic potential with final-state interactions. The data points are from SAID~\cite{GWU}.}
	\label{s11full}
\end{figure*}
\begin{table}[tb]
	\caption{The fit parameters obtained via the experimental data of the $S_{11}$ partial wave pion photoproduction. The $N^*_1$ and $N^*_2$ represent the bare states which mix to form the $N^*(1535)$ and $N^*(1650)$ resonances. Error estimates are obtained through the allowed variation in the $\Lambda_{-}^{\rm em}$.} 
	\centering
	\label{couplings11}
        \def\arraystretch{1.3}
  \begin{ruledtabular}
    \begin{tabular}{cr}

		$1/2^-$ Resonance Parameters &Value\\
		\hline
            $\Lambda_{-}^{\rm em}$&$1.007^{+0.150}_{-0.150}$\\
		$f_{\eta NN}$         &$-3.912^{+0.426}_{-0.619}$\\
		$g_{p\gamma N^{*+}_1}$&$1.049^{+0.019}_{-0.022}$\\
		$g_{n\gamma N^{*0}_1}$&$-0.349^{+0.011}_{-0.010}$\\
		$g_{p\gamma N^{*+}_2}$&$1.064^{+0.125}_{-0.095}$\\
		$g_{n\gamma N^{*0}_2}$&$-0.188^{+0.167}_{-0.198}$\\	

	\end{tabular}
  \end{ruledtabular}
\end{table}

The results for the $E_{0+}$ amplitude are shown in Fig.~\ref{s11full}, with the corresponding fit parameters listed in Table~\ref{couplings11}. The solid line in the figure represents our result, which was obtained by combining the electromagnetic transition potential from $\gamma N$ to the meson-baryon coupled channels considered in our model with the final-state interactions found in the earlier HEFT analysis.

In the electromagnetic transition process, we consider the contributions from two bare states which mix in forming the $N^*(1535)$ and $N^*(1650)$ resonances.  These single-particle bare states appear as intermediate propagators in the $s$- and $u$-channel Feynman diagrams. By setting the coupling constants $\hat{g}_{N\gamma N^*}$ to zero, we can evaluate the magnitude of their contributions, which are represented by the dashed lines in the figure. The coupling constants of pseudoscalar-octet mesons with two octet baryons $g_{PBB'}$ follow the SU(3) relations in EBAC, while the couplings $f_{\eta NN}$ and $f_{K\Sigma N}$ are allowed to vary. The former is varied in our model because we do not explicitly include the contribution from the $\eta'$ meson.

Within HEFT, the two low-lying odd-parity nucleon resonances, $N^*(1535)$ and $N^*(1650)$, are interpreted as quark-model like states, dressed by meson-baryon interactions~\cite{Abell:2023nex}. The dominant roles are played by the bare states with masses of 1.6301 GeV and 1.8612 GeV, respectively. The rescattering effects in the hadronic coupled channels lower the pole masses of these two resonances. 

When fitting the SAID data points, we exclude those in the 1.65 GeV to 1.70 GeV range. This is necessary because the SAID data points between 1.6 GeV and 1.7 GeV exhibit a bump in the imaginary part of the amplitude. This deviation occurs near the threshold of the coupled channel $K\Sigma$ channel, which may have significant coupling to these two resonances~\cite{Xie:2007qt, Doring:2008sv, Doring:2009uc, Li:2024rqb}. Incorporating the influence of this channel might further improve our results.

As mentioned earlier, the observed 
$N^*(1535)$ and $N^*(1650)$ resonances each have large bare state components. By setting the parameters $\hat{g}_{N\gamma N^*}$ to zero, we can clearly discern the significance of the contributions from these bare states, consistent with our analysis within HEFT. 
%%%%%%%RRRRRRRRRRRRRRR
The main contributions to the electromagnetic transition amplitudes  from the bare states occur through the $s$ channel. As the center-of-mass energy approaches the resonance region, the contributions from the bare states become more obvious, especially for the proton target. They provide a significant contribution to the electromagnetic potential, improving the consistency of our results with experimental data.

\begin{figure}[tb]
	\centering
	\begin{tabular}{cc}
		\includegraphics[width=7.5cm]{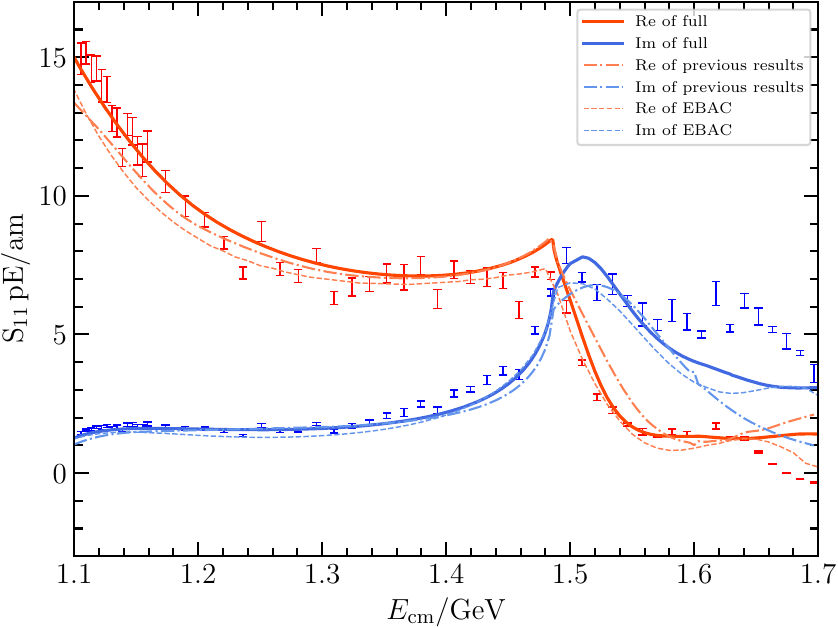}
	\end{tabular}
	\caption{The amplitudes $E_{0+}$ for a proton target from SAID~\cite{GWU} and EBAC~\cite{site:SL}, along with our previous results~\cite{Guo:2022hud} and the current calculation.
}
	\label{s11pourebac}
\end{figure}

We incorporate the contribution from the second bare state mixing to form the $N^*(1650)$, as well as the coupled $K\Lambda$ channel, in addition to the contributions included in our previous calculation~\cite{Guo:2022hud}. In that previous work, we used a Breit-Wigner propagator to estimate the influence of the $N^*(1650)$. Including this contribution slightly increased the imaginary part of the multipole amplitude in the 1.5 GeV-1.6 GeV range. Recently, these two nucleon resonances, $N^*(1535)$ and $N^*(1650)$,  have been jointly considered within HEFT. The coupling constants of the bare states with the coupled channels $\pi N$, $\eta N$ and $K\Lambda$ are obtained by fitting to the $S_{11}$ $\pi N$ scattering data and validated against the lattice QCD spectrum. In Fig.~\ref{s11pourebac}, we display the current results, along with our previous results
\footnote{
In our previous work~\cite{Guo:2022hud}, we had missed the Clebsch–Gordan coefficients for the $\rho$ exchange contributions of the $E_{0+}$ in the program code. For the $u$-channel diagram involving the bare state, $\hat g_{N\gamma N^*}$ should be replaced with $\frac{-\tau_3+3}{2}\frac{g_{p\gamma N^{*+}}}{3}+\frac{\tau_3+3}{2}\frac{g_{n\gamma N^{*0}}}{3}$ in Eq.~(15) of Ref.~\cite{Guo:2022hud}. These errors do not affect the main conclusions. We have corrected them in this study.
}
and those from EBAC and SAID for a  proton target. The imaginary parts of our results and those from EBAC are both smaller than the results from SAID in the 1.6 GeV to 1.7 GeV range. It can be observed that the results for the imaginary part become much better than our previous results above 1.5 GeV. 

%%%%%%%%%%%%%%%%%%%%%%%%%%%%%%%%%%%%%%%%%%%%%%%%%%%%%%%%%%%%%%%%%%%%%%%%%%%%%

\subsection{The $P_{11}$ partial wave amplitude}
In calculating the $P_{11}$ partial wave amplitude, we consider three coupled channels, $\pi N$, $\sigma N$ and $\pi\Delta$. Using the well-defined effective Lagrangians, we derive the electromagnetic transition potential from $\gamma N$ to these channels. The contributions from a possible bare state have also been considered. By applying HEFT to handle their final-state interactions, we obtain the total scattering T-matrix and the multipole amplitude $M_{1-}$. 

The parameters $g_{\sigma NN}$ was adjusted freely in the EBAC fit. Moreover, unlike $f_{\pi N\Delta}$, which can, in principle, be determined by the $\Delta\rightarrow\pi N$ process, $g_{\sigma NN}$ cannot be reliably fixed by any specific physical process. Therefore, We vary the parameter $g_{\sigma NN}$ freely in the analysis of the electromagnetic transition process. We introduce an exponential form factor into the electromagnetic potential.
%, which can enhance our fitting results. 

%
\begin{figure*}[htbp]
	\centering
	\begin{tabular}{cc}
		\includegraphics[height=6cm]{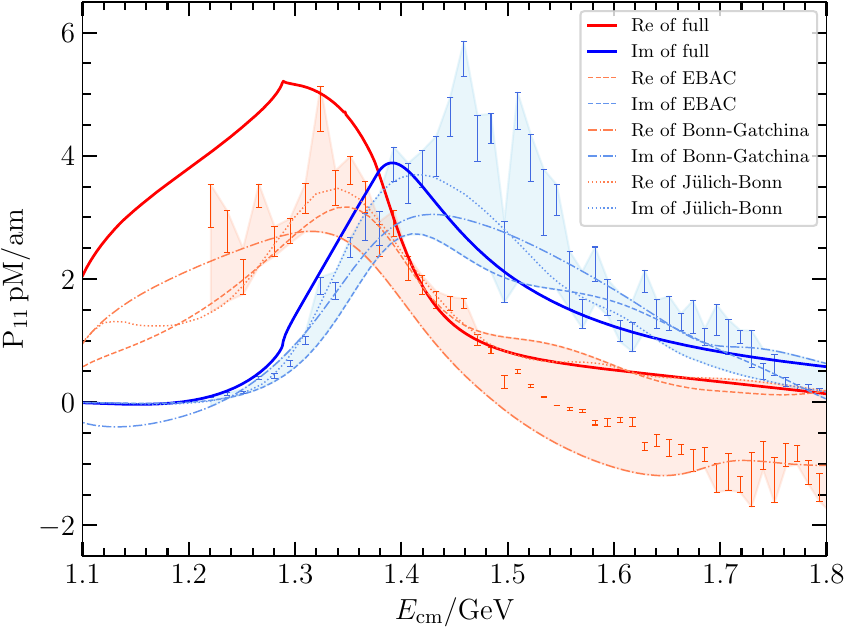}&
		\includegraphics[height=6cm]{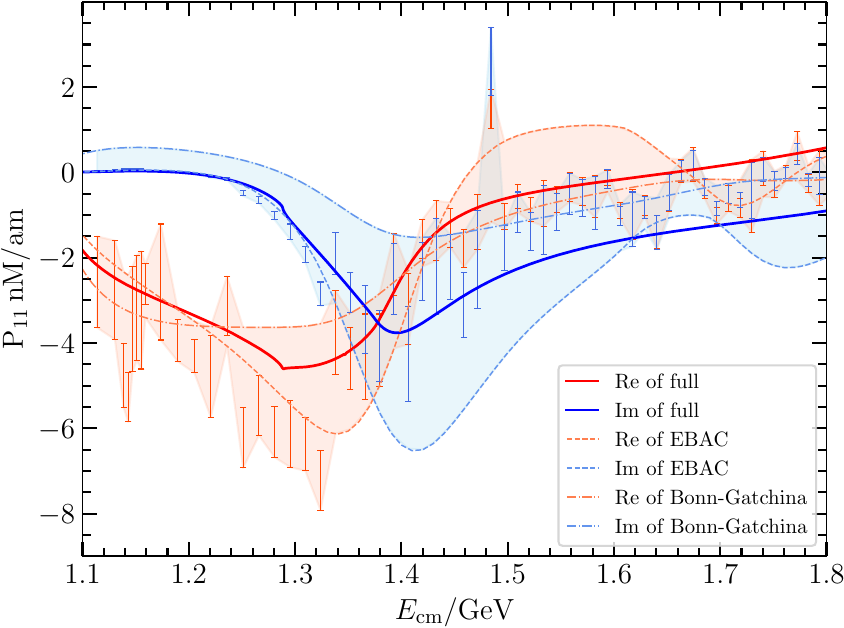}\\
		(a)&(b)
	\end{tabular}
	\caption{The fitted amplitude $M_{1-}$ from the present calculation, SAID~\cite{GWU} and other dynamical coupled channel models, along with the EBAC~\cite{site:SL}, Bonn-Gatchina Partial Wave Analysis~\cite{BG} and the analysis by the J{\"u}lich-Bonn-Washington Collaboration~\cite{jubo, Ronchen:2022hqk}. The shaded area represents the upper and lower limits of their results combined.}
	\label{p11compare}
\end{figure*}
\begin{table}[b]
	\caption{The fit parameters obtained from the analysis of the experimental data for pion photoproduction in the $P_{11}$ partial wave.}
	\centering
	\label{couplingp11}
        \def\arraystretch{1.3}
  \begin{ruledtabular}
    \begin{tabular}{cr}

		$1/2^+$ Resonance Parameters &Value\\
		\hline
            $\Lambda^{\rm em}_{+}$ (GeV)&$1.942^{+0.150}_{-0.150}$\\
		$g_{\sigma NN}$&$0.359^{+0.746}_{-0.000}$\\
		$g_{p\gamma N^{*+}}$&$0.323^{+0.358}_{-0.428}$\\
		$g_{n\gamma N^{*0}}$&$-1.132^{+0.477}_{-0.341}$\\
		$g_{\gamma\Delta^+N^{*+}}$&$-5.000^{+0.000}_{-0.000}$\\
		$g_{\gamma\Delta^0N^{*0}}$&$-5.000^{+0.000}_{-0.000}$\\	

	\end{tabular}
   \end{ruledtabular}
\end{table}

The results of this analysis are shown in Fig.~\ref{p11compare}, with the fit parameters listed in Table~\ref{couplingp11}. The solid line in the figure represents our calculated results. We also present the results of other dynamical coupled channel models. The SAID and Bonn-Gatchina models use the K-matrix approach to analyze pion- and photon-induced reactions~\cite{Workman:2012hx, Workman:2012jf, Anisovich:2011fc}. The SAID group employs the Chew-Mandelstam function to parameterize the hadronic scattering process~\cite{Workman:2012hx, Workman:2012jf}. In the Bonn-Gatchina model, background terms are added to the K-matrix to describe non-resonant transitions between hadronic channels~\cite{Anisovich:2011fc}. Both models use the P-vector approach to describe photon-induced reactions. The ANL-Osaka and Jülich-Bonn models are similar dynamical coupled channel models~\cite{Matsuyama:2006rp, Ronchen:2014cna}. By solving the coupled channel integral equations, they obtain the T-matrix between hadronic channels, which consists of both background and resonant parts. By constructing the photon vertex function, the photoproduction T-matrix can be evaluated. When constructing the potentials in the coupled channel equations, these two models exhibit some differences; for example, they include different exchanged particles and follow various constraints. The FSI effects were analyzed by using the HEFT in our approach. HEFT has been validated against lattice QCD simulations in addition to the scattering data, enhancing our understanding of the resonance structures and their dynamic generation mechanisms. There are also some differences in the potentials. The kernel $V_{\alpha, \gamma N}$ in Eq.~(\ref{finalT}) is derived by using effective Lagrangians, whereas the potentials involving the bare state with $\gamma N$ in the ANL-Osaka model are parameterized as a function with constants that need to be fitted~\cite{Kamano:2013iva}. In the Jülich-Bonn model these potentials are parameterized as polynomials~\cite{Ronchen:2014cna}.

As can be seen in Fig.~\ref{p11compare}, there are some differences between these models. We use the shaded area to represent the upper and lower limits of their results. We fit to the data points from SAID to obtain our results and the parameters in our model. The amplitude for the neutron target fits well, but there are some discrepancies between our results and the SAID data for the proton target. These experimental data have significant uncertainties. Additionally, the considerable differences between various models highlight the model dependence of the $P_{11}$ partial wave results. Our results generally lie within the range of these models.

The harmonic oscillator model, successful in explaining hadron spectroscopy, predicts the first positive-parity excited state (the $2s$ state) to lie in the region around 2 GeV. This is naively inconsistent with the experimental observation of the Roper resonance around 1.44 GeV. However, within HEFT the $N^*(1440)$ is described as the result of strong rescattering between the coupled meson-baryon channels, $\pi N$, $\sigma N$ and $\pi\Delta$ with a small quark-model-like state component having a mass near 2 GeV, in accord with lattice QCD~\cite{Wu:2017qve}. Although the bare mass of the $2s$ excitation is so high, the  strong rescattering effects provide a pole mass at approximately 1370 MeV. 

\begin{figure*}[htbp]
	\centering
	\begin{tabular}{cc}
		\includegraphics[height=6cm]{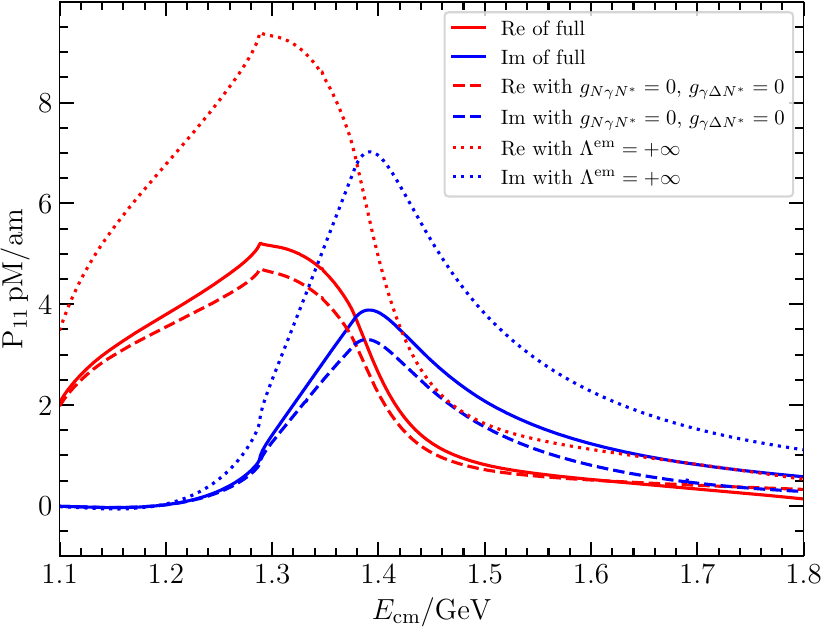}&
		\includegraphics[height=6cm]{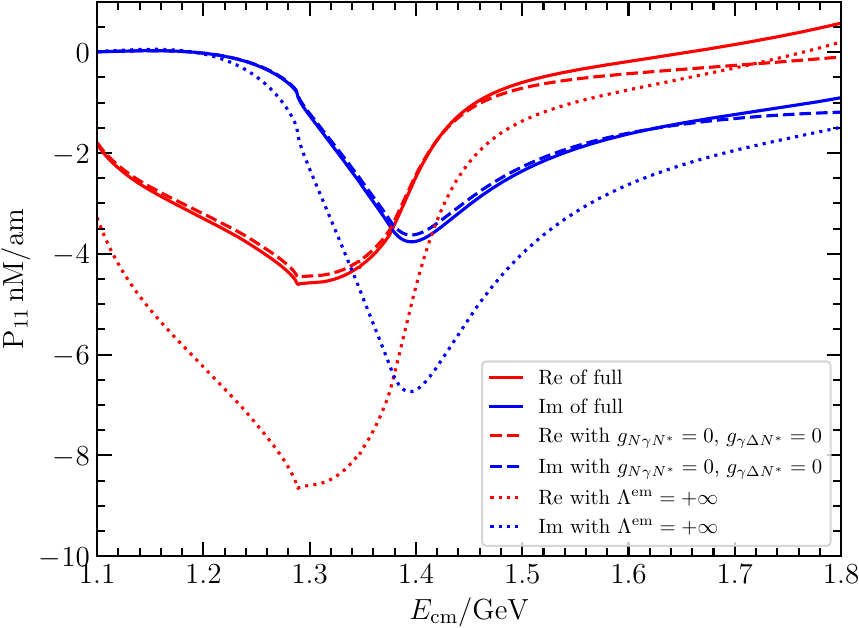}\\
		(a)&(b)
	\end{tabular}
	\caption{The fitted amplitude $M_{1-}$. The dashed line represents the results with the coupling constant $\hat{g}_{N\gamma N^*}$ and $\hat{g}_{\gamma\Delta N^*}$ set to zero. The dotted line shows the results without the form factor in the electromagnetic transition process.}
	\label{p11ff}
\end{figure*}

In Fig.~\ref{p11ff}, our complete results are represented by the solid line, while the results with the coupling constants $\hat{g}_{N\gamma N^*}$ and $\hat{g}_{\gamma\Delta N^*}$ set to zero are represented by the dashed line. Because of the large mass of the bare state, it does not contribute significantly to the $P_{11}$ partial wave within the energy range of our calculations.

When calculating the electromagnetic potential in the $P_{11}$ partial wave, we introduce an exponential form factor to suppress the contributions from the $\pi\Delta$ channel within the center-of-mass energy range of 1.2 GeV to 1.4 GeV. In Fig.~\ref{p11ff}, the dotted line represents the results without form factors in the electromagnetic process, achieved by setting the cutoff values $\Lambda^{\rm em}_{+}$ to infinity. Introducing these form factors significantly improves our results, particularly in the energy range of 1.2 GeV to 1.4 GeV. The contribution from the $\pi\Delta$ channel is the most substantial in the $P_{11}$ channel and would thus be affected the most by this regulator.

When calculating the multipole amplitude $E_{0+}$, the contact term,  $\hat{\mathcal{M}}_{cont}$, in the $S$-wave coupling gives the main contribution to the amplitude. In the calculation of $M_{1-}$, the leading-order contact terms in $\mathcal{M}_{\alpha,\gamma N}$ are independent of the angle $\theta$ in our approach, and therefore they vanish after integrating over $\theta$ for P-wave channels. The primary contribution in this case comes from the $s$-channel Feynman diagram of the $\pi\Delta$ coupled channel through exchanging nucleons. This differs from the calculation of $E_{0+}$, where the reaction parity is negative. To ensure parity conservation, only the anti-nucleon can propagate in the $s$-channel, thereby reducing its contribution. In contrast, for the calculation of $M_{1-}$, the reaction parity is positive, allowing the nucleon to propagate in the $s$-channel, resulting in a larger contribution. As the center-of-mass energy increases, the contribution from the $\pi\Delta$ coupled channel gradually grows, peaking around 1.3 GeV to 1.4 GeV. When the energy exceeds 1.4 GeV, this contribution significantly decreases, becoming comparable to other contributions, which causes the multipole amplitude to noticeably diminish.

Additionally, we refit $g_{\sigma NN}$ to a new value of 0.359, indicating that the contributions from the $\sigma N$ coupled channel are suppressed in our results compared to those from EBAC. In Scenario I of our previous work on the $ N^*(1440) $~\cite{Wu:2017qve}, the coupling of the bare state to $ \pi N $ and $ \sigma N $ is suppressed compared to Scenario II, indicating that the $ N^*(1440) $ is generated primarily by strong rescattering effects. The coupling constant $ g_{B_0\sigma N} $ in this Scenario is set to zero, which suppresses the contributions of the $ \sigma N $ channel.

Because the $\Delta$ is unstable, we should include a complex mass shift in the $\pi\Delta$ propagator in Eq.~(\ref{finalT}). In the EBAC analysis, the mass shift term involving $\pi\Delta$ takes the following form~\cite{Matsuyama:2006rp, Kamano:2013iva}
\begin{align}
	\Sigma_{\pi\Delta}(k;E_{\mathrm{cm}})=&\frac{m_\Delta}{\omega_\Delta(k)}\int q^2 dq\frac{\omega_{\pi N}(q)}{\left[\omega_{\pi N}^2(q)+k^2\right]^{1/2}}\notag\\
	&\times\frac{|f_{\Delta\to\pi N}(q)|^2}{E_{\mathrm{cm}}-\omega_\pi(k)-\left[\omega_{\pi N}^2(q)+k^2\right]^{1/2}+i\epsilon} \, ,
\end{align}
with
\begin{align}
	f_{\Delta\to\pi N}(q)=-i\frac{0.98}{[2(m_N+m_\pi)]^{1/2}}\frac{q}{m_\pi}\left(\frac{1}{1+[q/(0.358 \mathrm{GeV})]^2}\right)^2 \, .
\end{align}
\begin{figure}[htbp]
	\centering
	\begin{tabular}{cc}
		\includegraphics[height=5.5cm]{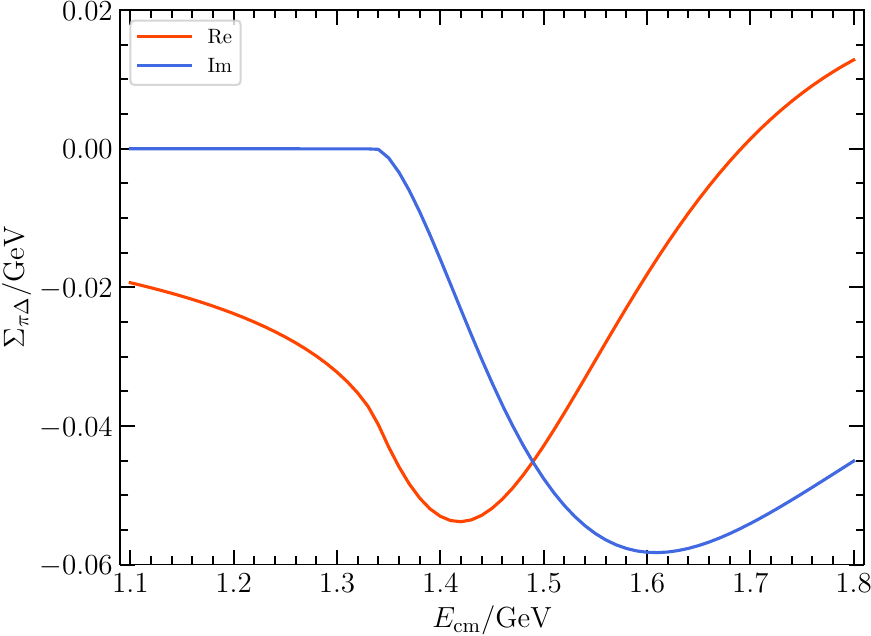}
	\end{tabular}
	\caption{The variation of $\Sigma_{\pi\Delta}(k;E_{\rm cm})$ with the center-of-mass energy.}
	\label{sigma}
\end{figure}
\begin{figure}[htbp]
	\centering
	\begin{tabular}{cc}
		\includegraphics[height=5.5cm]{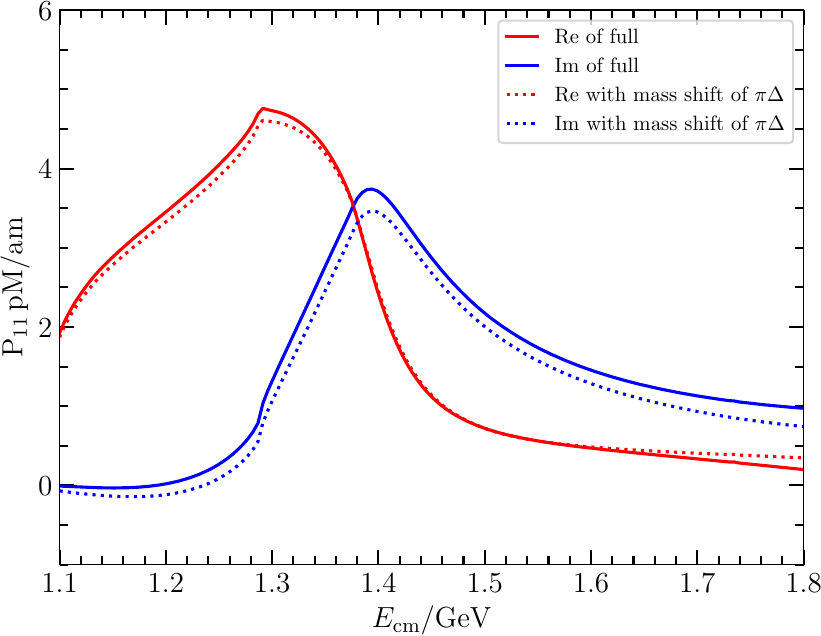}
	\end{tabular}
	\caption{The amplitude $M_{1-}$ with and without the contribution from the $\pi\Delta$ mass shift for the proton target.}
	\label{p11im}
\end{figure}

Here $k$ is the center-of-mass three-momentum in the $\pi\Delta$ channel. When $k=0.2$ GeV, the variation of $\Sigma_{\pi\Delta}(k;E_{\rm cm})$ with the center-of-mass energy is illustrated in Fig.~\ref{sigma}. As the center-of-mass energy increases, both the real and imaginary parts of the mass shift term initially increase and then decrease. The maximum value of the imaginary part is approximately half the width of $\Delta$. To estimate the contribution from the mass shift term $\Sigma_{\pi\Delta}(k;E_{\rm cm})$, we can simply set $\Sigma_{\pi\Delta}(k;E_{\rm cm})=-0.05-0.06i$ GeV, which is nearly the maximum value shown in Fig.~\ref{sigma}. As shown in Fig.~\ref{p11im}, even when the mass shift term is set to its maximum value in this case, the contribution from introducing the mass shift effect in the $\pi\Delta$ propagator is relatively small.

Finite-volume HEFT is designed to reproduce the leading behavior of finite-volume ChPT in the perturbative limit~\cite{Hall:2013qba, Abell:2021awi}. However, the study of resonance phenomena and its associated avoided level crossings in finite volume require a non-perturbative effective field theory (EFT) formalism. Thus, HEFT may be regarded as a non-perturbative extension founded on ChPT. Model independence is maintained only if the Hamiltonian accurately describes experimental data and lattice QCD results.

In reproducing the leading behavior of finite-volume ChPT in the perturbative limit HEFT is formulated with the leading interactions of the chiral Lagrangian. In principle, higher-order terms cannot be neglected in the non-perturbative solution, but at the current level of accuracy, the leading terms are sufficient to describe the scattering data in our framework. One may need to include higher-order terms as the accuracy of scattering data and lattice QCD results improve.

With this understanding, one can commence with the consideration of uncertainties in the FSI.  One can adjust the regulators of HEFT, repeating the fits to the experimental scattering data and lattice QCD results and examining the extent to which HEFT can still describe these data in an accurate manner.  This approach was recently examined in Ref.~\cite{Liu:2023xvy} where the uncertainties were found to be very small due to the demands of an accurate description of the data. The regulator parameters were constrained by data and lattice QCD to a very narrow range of 0.9 to 1.1 GeV, a variation of $\pm100$ MeV.  Given that the Hamiltonians for the final-state effects are tightly constrained, one can proceed in the same spirit and change the regulator parameters in the electromagnetic sector, refitting the experimental data and examining the variation in the results.

\begin{figure*}[htbp]
	\centering
	\begin{tabular}{cc}
		\includegraphics[height=6cm]{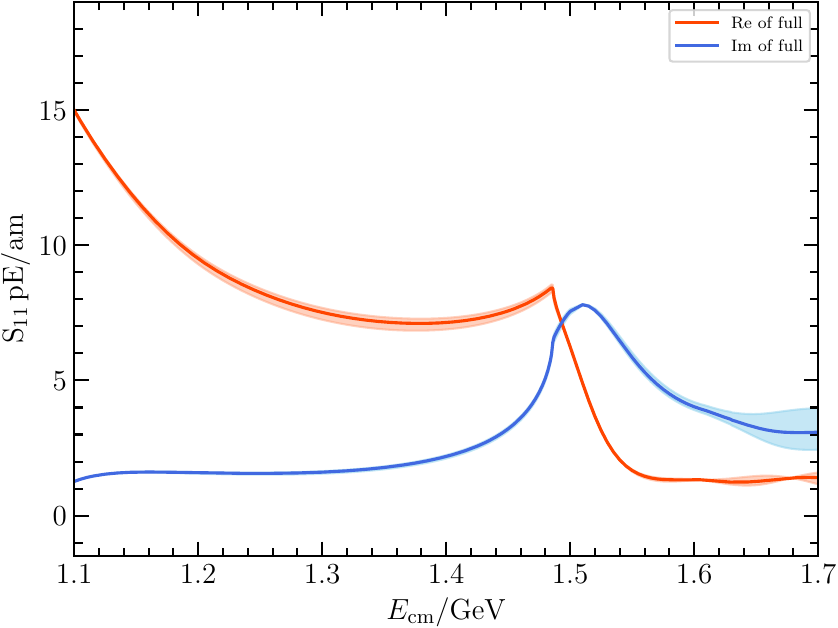}&
		\includegraphics[height=6cm]{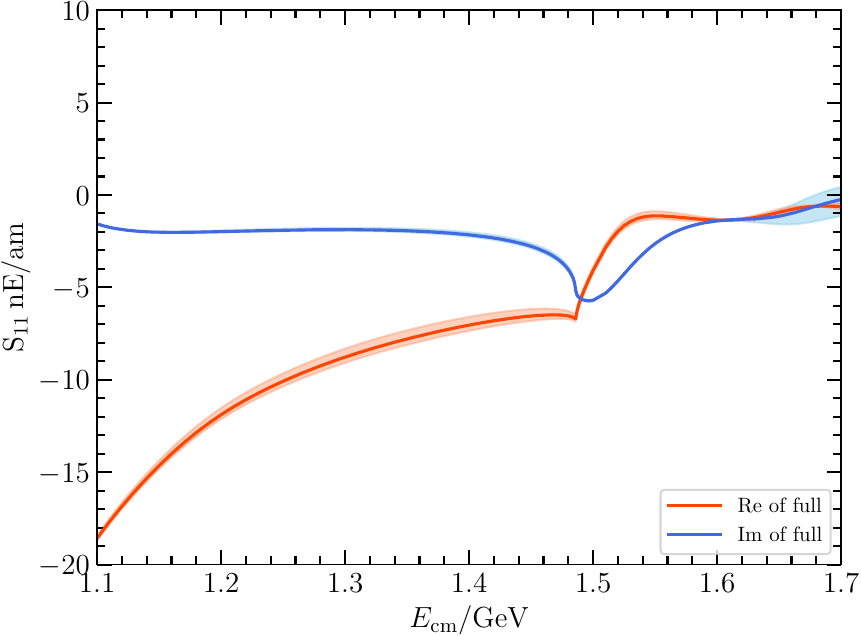}\\
		(a)&(b)
	\end{tabular}
	\caption{Uncertainty estimate for the $E_{0+}$ amplitude. To explore alternative descriptions, the regulator parameter $\Lambda^{\rm em}_{-}$ is varied by $\pm150$ MeV as the coupling parameters are adjusted to describe the experimental data.}
	\label{errorS11}
\end{figure*}

\begin{figure*}[htbp]
	\centering
	\begin{tabular}{cc}

  \includegraphics[height=6cm]{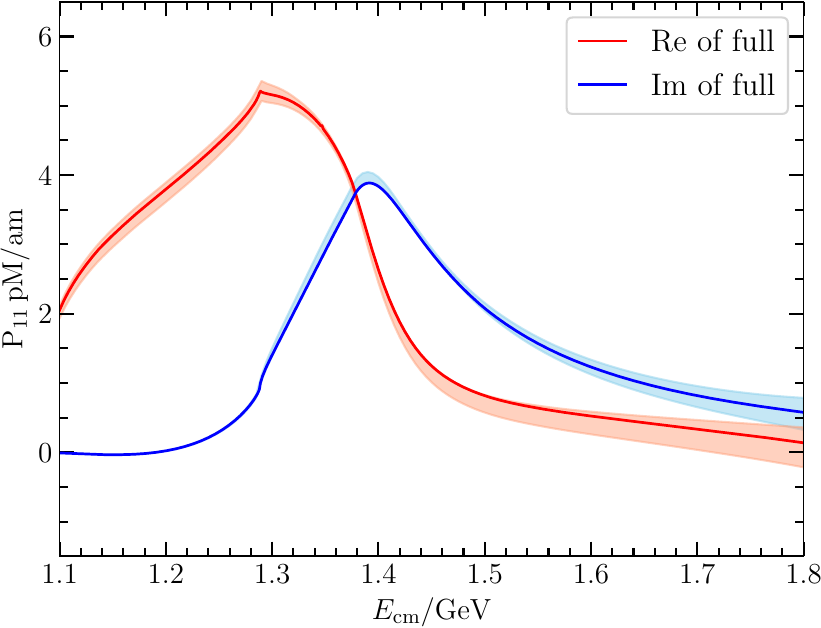}&
		\includegraphics[height=6cm]{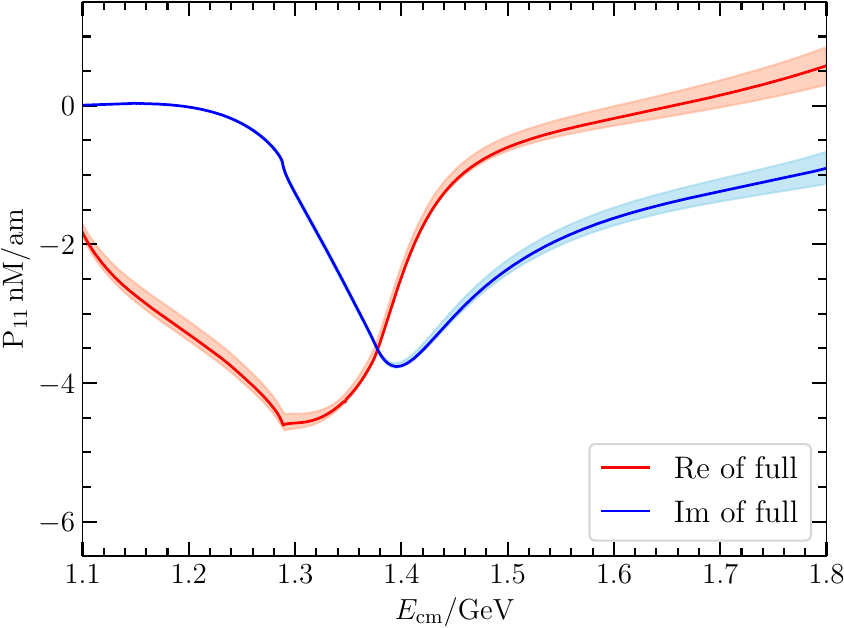}\\
		(a)&(b)
	\end{tabular}
	\caption{Uncertainty estimate for the $M_{1-}$ amplitude. To explore alternative descriptions, the regulator parameter $\Lambda^{\rm em}_{+}$ is varied by $\pm150$ MeV as the coupling parameters are adjusted to describe the experimental data.}
	\label{errorP11}
\end{figure*}

To explore the uncertainty, the regulator parameters of the form factors can be varied and the coupling parameter redetermined by fitting the experimental data. This time we vary the regulator parameter more generously, varying the regulator $\Lambda^{\rm em}$ by $\pm150$ MeV and refitting the parameters. The fitting result are shown in Fig.~\ref{errorS11} and Fig.~\ref{errorP11}. The variation in the form factor regulator is largely compensated through variation of the coupling parameters. Similar to the findings of Ref.~\cite{Liu:2023xvy}, we observe the accuracy of the experimental data governs the uncertainty in our results. 

%

%%%%%%%%%%%%%%%%%%%%%%%%%%%%%%%%%%%%%%%%%%%%%%%%%%%%%%%%%%%%%%%%%%%%%%%%%%%%%

\section{Summary}\label{summary}
Pion photoproduction is an essential reaction for identifying the existence of baryon excited states and extracting the properties of these resonances. 

We have refined our calculation of the electric multipole amplitude $E_{0+}$ in the $S_{11}$ partial wave for $\gamma N\rightarrow\pi N$ by including the contributions from the $N^*(1650)$ as well as the $K\Lambda$ coupled channel. The T-matrix for $\gamma N\rightarrow\pi N$ is split into two parts. The first, involving the photon absorption, was derived from well-defined effective Lagrangians. In this part of the calculation we also considered contributions from the two bare states associated with the $N^*(1535)$ and $N^*(1650)$. 

The second piece of the calculation involved the final-state interactions, which  were studied in previous work, using Hamiltonian Effective Field Theory. The parameters obtained by fitting to experimental data in that earlier work were not altered here. This considerably simplified our calculation. By including only the coupled channels whose thresholds lie within the energy range considered here, we have obtained results consistent with experimental data. The numerical results for $E_{0+}$ show that the contributions from the bare states are indispensable, which is consistent with the conclusion within HEFT that these two low-lying odd-parity nucleon resonances contain a significant bare state component~\cite{Abell:2023nex}.

We also calculated the magnetic multipole amplitude, $M_{1-}$, in the $P_{11}$ partial wave to gain more insight into the nature of the $N^*(1440)$. Within HEFT, the 
$N^*(1440)$ is a dynamically generated resonance, with a small bare state component having a mass of order 
2 GeV~\cite{Wu:2017qve}. The large mass of the bare state results in a relatively small contribution to the overall amplitude. The experimental values from SAID have significant uncertainties. While the results from different dynamical coupled-channel models show considerable variation, our results generally fall between them. To obtain more accurate physical results and reduce model dependence, it would be helpful to further improve the experimental precision in the future.

\iffalse
It is important to recognise the considerable interest that has been shown in the transition form factors (or helicity amplitudes)~\cite{Aznauryan:2004jd,Burkert:2017djo}, especially to the Roper resonance, derived from the experimental data. It should be clear that,  because the present calculation reproduces the photoproduction data, it is consistent with those derived transition form factors, even though here the Roper is dynamically generated.
\fi

The three-particle $\pi\pi N$ channel is an important decay mode of $N^*(1440)$~\cite{Workman:2022ynf}. Using HEFT, incorporating the $\pi\pi N$ coupled channel in future studies of $N^*(1440)$ may help us better understand its structure and dynamical generation mechanisms.

%%%%%%%%%%%%%%%%%%%%%%%%%%%%%%%%%%%%%%%%%%%%%%%%%%%%%%%%%%%%%%%%%%%%%%%%%%%%%%%%%%%

\section{Acknowledgment}
We would like to thank Curtis Abell and Dan Guo for useful discussions. This work was supported by the National Natural Science Foundation of China under Grant Nos. 12175091, 12335001, and 12247101, the 111 Project under Grant No. B20063, and the innovation project for young science and technology talents of Lanzhou city under Grant No. 2023-QN-107. This research was supported by the University of Adelaide and by the Australian Research Council through ARC Discovery Project Grants No. DP190102215 and No. DP210103706 (D.B.L) and No. DP230101791 (A.W.T).

\appendix
\section{Lagrangians and amplitudes}\label{app1}
The effective Lagrangians~\cite{Matsuyama:2006rp, Kamano:2013iva} we use and the electromagnetic transition amplitudes $\hat{\mathcal{M}}_{\alpha,\gamma N}$ defined in Eq.~(\ref{pnM}) are listed below.
%%%%%%%%%%%%%%%%%%%%%%%%%%%%%%
\subsection{Odd parity case}
The $N^*(1535)$ and $N^*(1650)$ both have an odd parity. We consider three coupled channels, $\pi N$, $\eta N$ and $K \Lambda$. The effective Lagrangians involved are as follows
\begin{gather}
	\mathcal{L}_{\pi NN}=-\frac{f_{\pi NN}}{m_\pi}\bar{N}\gamma_\mu\gamma_5 \vec{\tau} \cdot \partial^\mu\vec{\pi} N \, ,
\end{gather}
\begin{gather}
    \mathcal{L}_{\gamma NN}=e\bar{N}\left[ \hat{e}_N\gamma^\mu A_\mu + \frac{\hat{\kappa}_N}{4m_N}\sigma^{\mu\nu} F_{\mu\nu}\right] N \, ,
\end{gather}
\begin{gather}
    \mathcal{L}_{\gamma\pi\pi}=e\left[ \vec{\pi}\times\partial^\mu\vec{\pi} \right]_3 A_\mu \, ,
\end{gather}
\begin{gather}
    \mathcal{L}_{\gamma N\pi N}=e\frac{f_{\pi NN}}{m_\pi}\bar{N} \gamma^\mu\gamma_5 \left[ \vec{\tau}\,\times\vec\pi \right]_3 N A_\mu \, ,
\end{gather}
\begin{gather}
    \mathcal{L}_{\gamma \rho\pi}=e\frac{g_{\gamma \rho\pi}}{m_\pi}\varepsilon^{\mu\nu\alpha\beta} \vec{\pi}\cdot \partial_\mu\vec{\rho}_\nu\partial_\alpha A_\beta \, ,
\end{gather}
\begin{gather}
	\mathcal{L}_{\rho NN}=g_{\rho NN}\bar{N} \left[ \gamma_\mu - \frac{\kappa_\rho}{2m_N} \sigma_{\mu\nu}\partial^\nu\right]\vec{\rho}\,^\mu\cdot\frac{\vec{\tau}}{2} N \, ,
\end{gather}
\begin{gather}
    \mathcal{L}_{\eta NN}=-\frac{f_{\eta NN}}{m_\eta}\bar{N} \gamma_\mu\gamma_5 N \partial^\mu \eta \, ,
\end{gather}	
\begin{gather}
    \mathcal{L}_{\gamma\rho\eta}=e\frac{g_{\gamma\rho\eta}}{m_\rho}\varepsilon^{\mu\nu\alpha\beta} \partial_\mu\rho^0_\nu\partial_\alpha A_\beta \eta \, ,
\end{gather}
\begin{gather}
    \mathcal{L}_{KN\Lambda}=-\frac{f_{KN\Lambda}}{m_K}\bar{\Lambda}\gamma^{\mu}\gamma_5\partial_\mu \bar{K}N+{\rm H.c.} \, ,
\end{gather}
\begin{gather}
	\mathcal{L}_{KN\Sigma}=-\frac{f_{KN\Sigma}}{m_K}\gamma^{\mu}\gamma_5(\partial_\mu \bar{K}\bar{\vec{\Sigma}}\cdot\vec{\tau}N)+{\rm H.c.} \, ,
\end{gather}
\begin{gather}
    \mathcal{L}_{\gamma\Lambda\Lambda}=e\bar{\Lambda}\frac{\kappa_\Lambda}{4m_N}\sigma^{\mu\nu}F_{\mu\nu}\Lambda \, ,
\end{gather}
\begin{gather}
    \mathcal{L}_{\gamma\Lambda\Sigma}=e\bar{\Lambda}\frac{\kappa_{\Lambda\Sigma}}{4m_N}\sigma^{\mu\nu}F_{\mu\nu}\Sigma+{\rm H.c.} \, ,
\end{gather}
\begin{gather}
    \mathcal{L}_{\gamma NK\Lambda}=ie\frac{f_{KN\Lambda}}{m_K}\left[\bar{p}\slashed{A}\gamma_5K^+\Lambda-\bar{\Lambda}\slashed{A}\gamma_5K^-p\right] \, ,
\end{gather}
\begin{gather}
    \mathcal{L}_{\gamma KK}=ie\left[K^-\partial^\mu K^+-(\partial^\mu K^-)K^+\right]A_\mu \, ,
\end{gather}
\begin{gather}
    \mathcal{L}_{\gamma K^*K}=e\frac{g^0_{\gamma K^*K}}{m_K}\epsilon_{\alpha\beta\gamma\delta}\left[\bar{K}^0(\partial^\gamma K^{*0,\delta})+K^0(\partial^\gamma\bar{K}^{*0,\delta})\right]\partial^\alpha A^\beta\notag\\
	+e\frac{g^c_{\gamma K^*K}}{m_K}\epsilon_{\alpha\beta\gamma\delta}\left[K^+(\partial^\gamma K^{*-,\delta})+K^-(\partial^\gamma K^{*+,\delta})\right]\partial^\alpha A^\beta \, ,
\end{gather}
\begin{gather}
    \mathcal{L}^{1/2\pm}_{\gamma NN^*}=e\bar N^*  \frac{\hat{g}_{N\gamma N^*}}{4m_N}\Gamma^\pm \sigma_{\mu\nu}F^{\mu\nu} N +\mathrm{H.c.} \, ,
\end{gather}
\begin{gather}
	\mathcal{L}^{1/2\pm}_{\pi NN^*}=-\frac{g_{\pi NN^*}}{f_\pi} \bar N^*\gamma^\mu\Gamma^\mp\vec{\tau}\cdot\partial_\mu\vec\pi N + \mathrm{H.c.} \, ,
\end{gather}		
\begin{gather}
    \mathcal{L}^{1/2\pm}_{\eta NN^*}=\frac{\sqrt{3}g_{\eta NN^*}}{f_\pi} \bar{N}^*\gamma^\mu\Gamma^\mp\partial_\mu\eta N + \mathrm{H.c.} \, ,
\end{gather}
\begin{gather}
    \mathcal{L}^{1/2\pm}_{K\Lambda N^*}= \frac{\sqrt{3}g_{K\Lambda N^*}}{f_\pi} \bar{N}^*\gamma^\mu\Gamma^\mp\partial_\mu K\Lambda + \mathrm{H.c.} \, ,
\end{gather}		
where $\Gamma^+=I_{4\times 4}$, $\Gamma^-=\gamma_5$, $\hat{e}_N={\rm diag}\{+1, 0\}$, $\hat{\kappa}_N={\rm diag}\{\kappa_p=1.79, \kappa_n=-1.91\}$, $\kappa_\rho=1.82$, $\kappa_\Lambda=-0.61$, $\kappa_{\Lambda\Sigma}=-1.61$ and $\hat{g}_{N\gamma N^*}={\rm diag}\{g_{p\gamma N^{*+}},g_{n\gamma N^{*0}}\}$. The corresponding Feynman amplitude $\hat{\mathcal{M}}_{\alpha,\gamma N}$ is as follows

$\hat{\mathcal{M}}_{\pi N,\gamma N}$:
\begin{gather}
		\mathcal{\hat{M}}_{sN}=\frac{\sqrt3 e}{m_\pi}f_{\pi NN}\slashed{k}\gamma_5 \frac{1}{\slashed{q}+\slashed{p}-m_N}\Gamma_N u^{\rm em}_{-}(k) \, ,\label{pin1}
\end{gather}
\begin{align}
    \mathcal{\hat{M}}_{uN}=&\left(\frac{-\tau_3+3}{2}\frac{\sqrt3 e}{3m_\pi}f_{\pi NN}\Gamma_p\frac{1}{\slashed{p}-\slashed{k}-m_N} \slashed{k}\gamma_5\right.\notag\\
    &\left.+\frac{\tau_3+3}{2}\frac{\sqrt3 e}{3m_\pi}f_{\pi NN}\Gamma_n\frac{1}{\slashed{p}-\slashed{k}-m_N} \slashed{k}\gamma_5\right) u^{\rm em}_{-}(k)\, ,\label{pin2}
\end{align}
\begin{gather}
    \mathcal{\hat{M}}_{cont}=-\frac{2\sqrt3 e}{3m_\pi}f_{\pi NN}\tau^3\slashed{\varepsilon}_\gamma \gamma_5 u^{\rm em}_{-}(k)\, ,\label{pin3}
\end{gather}
\begin{gather}
    \mathcal{\hat{M}}_{t\pi}=\frac{2\sqrt3 e}{3m_\pi}f_{\pi NN}\tau^3 \frac{\tilde{\slashed{k}}\gamma_5}{\tilde{k}^2-m^2_\pi} (\tilde{k}+k) \cdot \varepsilon_\gamma u^{\rm em}_{-}(k)\, ,\label{pin4}
\end{gather}	
\begin{gather}
    \mathcal{\hat{M}}_{t\rho}=\frac{\sqrt3 e}{m_\pi}\frac{g_{\rho NN}g_{\gamma\rho\pi}}{2}
	\frac{\Gamma_\rho}{\tilde{k}^2-m^2_\rho} u^{\rm em}_{-}(k)\, ,\label{pin5}
\end{gather}	
\begin{align}\label{s11barep}
    \mathcal{\hat{M}}^{1/2-}_{s+u,N^*}=&e \frac{g_{\pi NN^*}}{f_\pi} \frac{\sqrt{3}}{4m_N}
	\left\{\hat{g}_{N\gamma N^{*}}\slashed{k} \frac{1}{\slashed{q}+\slashed{p}-m_{N^*}^0}  \gamma_5(\slashed{q}\slashed{\varepsilon}_\gamma- \slashed{\varepsilon}_\gamma\slashed{q})
	\right.\notag\\
	&+(\frac{-\tau_3+3}{2}\frac{g_{p\gamma N^{*+}}}{3}+\frac{\tau_3+3}{2}\frac{g_{n\gamma N^{*0}}}{3})\gamma_5(\slashed{q}\slashed{\varepsilon}_\gamma- \slashed{\varepsilon}_\gamma\slashed{q})\notag\\
	&\times\left.\frac{1}{\slashed{p}-\slashed{k}-m_{N^*}^0} \slashed{k} \right\}u^-(k) \, ,
\end{align}			

$\hat{\mathcal{M}}_{\eta N,\gamma N}$:
\begin{align}
	\mathcal{\hat{M}}_{sN}=-e\frac{f_{\eta NN}}{m_\eta}\slashed{k}\gamma_5 \frac{1}{\slashed{q}+\slashed{p}-m_N}\Gamma_N u^{\rm em}_{-}(k)\, ,
\end{align}
\begin{align}
    \mathcal{\hat{M}}_{uN}=-e\frac{f_{\eta NN}}{m_\eta}\Gamma_N\frac{1}{\slashed{p}-\slashed{k}-m_N} \slashed{k}\gamma_5 u^{\rm em}_{-}(k)\, ,
\end{align}
\begin{align}
    \mathcal{\hat{M}}_{t\rho}=- \,e\frac{g_{\rho NN}g_{\gamma\rho\eta}}{m_\rho} \frac{\tau^3}{2}    \frac{\Gamma_\rho}{\tilde{k}^2-m^2_\rho} u^{\rm em}_{-}(k)\, ,
\end{align}
\begin{align}
    \mathcal{\hat{M}}^{1/2-}_{s+u,N^*}={}&e \frac{\sqrt{3} g_{\eta NN^*}}{f_\pi} \frac{\hat{g}_{N\gamma N^*}}{4m_N}
	\left\{ \slashed{k} \frac{1}{\slashed{q}+\slashed{p}-m_{N^*}^0}  \gamma_5(\slashed{q}\slashed{\varepsilon}_\gamma- \slashed{\varepsilon}_\gamma\slashed{q})\right.\notag\\
	&\left.+\gamma_5(\slashed{q}\slashed{\varepsilon}_\gamma- \slashed{\varepsilon}_\gamma\slashed{q}) \frac{1}{\slashed{p}-\slashed{k}-m_{N^*}^0} \slashed{k} \right\} u^-(k) \, ,
\end{align}

$\hat{\mathcal{M}}_{K\Lambda,\gamma N}$:
\begin{align}
    \mathcal{\hat{M}}_{sN}=-e\frac{f_{KN\Lambda}}{m_{K}}\slashed{k}\gamma_{5}\frac{1}{\slashed{q}+\slashed{p}-m_N}\Gamma_{N} u^{\rm em}_{-}(k)\, ,
\end{align}
\begin{align}
    \mathcal{\hat{M}}_{u\Lambda}=-e\frac{f_{KN\Lambda}}{m_{K}}\frac{\kappa_\Lambda}{4m_N}(\slashed{q}\slashed{\varepsilon}_\gamma - \slashed{\varepsilon}_\gamma\slashed{q})\frac{1}{\slashed{p}-\slashed{k}-m_\Lambda}\slashed{k}\gamma_{5} u^{\rm em}_{-}(k)\, ,
\end{align}
\begin{align}
    \mathcal{\hat{M}}_{u\Sigma}=-e\frac{f_{KN\Sigma}}{m_{K}}\frac{\kappa_{\Lambda\Sigma}}{4m_N}\tau^3(\slashed{q}\slashed{\varepsilon}_\gamma - \slashed{\varepsilon}_\gamma\slashed{q})\frac{1}{\slashed{p}-\slashed{k}-m_\Sigma}\slashed{k}\gamma_{5} u^{\rm em}_{-}(k)\, ,
\end{align}
\begin{align}
	\mathcal{\hat{M}}_{cont}=e\frac{f_{KN\Lambda}}{m_K}\frac{\tau^3+1}{2}\slashed{\varepsilon}_\gamma\gamma_5 u^{\rm em}_{-}(k)\, ,
\end{align}		
\begin{align}
	\mathcal{\hat{M}}_{tK}=-e\frac{f_{KN\Lambda}}{m_{K}}\frac{\tau^3+1}{2}\frac{\slashed{\tilde{k}}\gamma_{5}}{\tilde{k}^{2}-m_{K}^{2}}(\tilde{k}+k^{\prime})\cdot\varepsilon_{\gamma} u^{\rm em}_{-}(k)\, ,
\end{align}		
\begin{align}	
	\mathcal{\hat{M}}_{tK^*}&=-ie\hat{g}_{\gamma KK^*}\frac{g_{K^{*}N\Lambda}}{m_{K}}\left[\gamma^{\delta}+\frac{\kappa_{K^{*}N\Lambda}}{2(m_{N}+m_{\Lambda})}(\gamma^{\delta}\slashed{\tilde{k}}-\slashed{\tilde{k}}\gamma^{\delta})\right]\notag\\
	&\times\epsilon_{\alpha\beta\eta\delta}\tilde{k}^{\eta}q^{\alpha}\varepsilon_{\gamma}^{\beta}\frac{1}{\tilde{k}^{2}-m_{K^{*}}^{2}} u^{\rm em}_{-}(k)\, ,
\end{align}			
\begin{align}
	\mathcal{\hat{M}}^{1/2-}_{s,N^*}=e \frac{\sqrt{3} g_{K\Lambda N^*}}{f_\pi} \frac{\hat{g}_{N\gamma N^*}}{4m_N}
	\slashed{k} \frac{1}{\slashed{q}+\slashed{p}-m_{N^*}^0}  \gamma_5(\slashed{q}\slashed{\varepsilon}_\gamma- \slashed{\varepsilon}_\gamma\slashed{q})u^-(k) \, ,
\end{align}
where $\tilde{k}=p-p^\prime$, $\Gamma_N=\hat{e}_N\slashed{\varepsilon}_\gamma + \frac{\hat{\kappa}_N}{4m_N} [\slashed{q}\slashed{\varepsilon}_\gamma - \slashed{\varepsilon}_\gamma\slashed{q}]$,
$\Gamma_\rho=i\varepsilon_{\mu\nu\alpha\beta}\tilde{k}^\mu[\gamma^\nu + \frac{\kappa_\rho}{4m_N}(\gamma^\nu\tilde{\slashed{k}}-\tilde{\slashed{k}}\gamma^\nu)] q^\alpha \varepsilon^\beta_\gamma$ and $\hat{g}_{\gamma KK^*}={\rm diag}\{g^c_{\gamma KK^*},\, g^0_{\gamma KK^*}\}$.
%%%%%%%%%%%%%%%%%%%%%%%%%%%%%%%%%%%%%%%%%%%%%%

\subsection{Even parity case}
The $N^*(1440)$ has an even parity. We consider three coupled channels, $\pi N$, $\sigma N$ and $\pi\Delta$. The effective Lagrangians~\cite{Matsuyama:2006rp, Kamano:2013iva} involved are as follows
\begin{gather}
    \mathcal{L}_{\sigma NN}=g_{\sigma NN}\bar{N}N\sigma \, ,
\end{gather}
\begin{gather}
    \mathcal{L}_{\pi N\Delta}=-\frac{f_{\pi N\Delta}}{m_\pi}\bar{\Delta}^\mu\vec{T}\cdot\partial_\mu\vec{\pi}N+{\rm H.c.} \, ,
\end{gather}
\begin{gather}
    \mathcal{L}_{\gamma N\Delta}=-ie\bar{\Delta}^\mu\Gamma^{\rm em,\Delta N}_{\mu\nu}T^3NA^\nu+{\rm H.c.} \, ,
\end{gather}
\begin{gather}
	\mathcal{L}_{\pi\Delta\Delta}=\frac{f_{\pi\Delta\Delta}}{m_{\pi}}\bar{\Delta}_\mu\gamma^\nu\gamma_5\vec{T}_{\Delta}\cdot\partial_\nu\vec{\pi}\Delta^\mu \, ,
\end{gather}
\begin{align}
	\mathcal{L}_{\gamma\Delta\Delta}=&e\bar{\Delta}^{\eta}\left(T_{\Delta}^{3}+\frac{1}{2}\right)\left[-\gamma_{\mu}g_{\eta\nu}+\big(g_{\mu\eta}\gamma_{\nu}+g_{\mu\nu}\gamma_{\eta}\big)\right.\notag\\
	&\left.+\frac{1}{3}\gamma_{\eta}\gamma_{\mu}\gamma_{\nu}\right]\Delta^{\nu}A^{\mu} \, ,
\end{align}
\begin{gather}
	\mathcal{L}_{\gamma\pi N\Delta}=e\frac{f_{\pi N\Delta}}{m_{\pi}}\left[(\bar{\Delta}^{\mu}\vec{T}N)\times\vec{\pi}\right]_{3}A_{\mu}+\mathrm{H.c.} \, ,
\end{gather}
\begin{gather}
	\mathcal{L}_{\rho N\Delta}=-i\frac{f_{\rho N\Delta}}{m_{\rho}}\bar{\Delta}^{\mu}\gamma^{\nu}\gamma_{5}\left[\partial_{\mu}\vec{\rho}_{\nu}-\partial_{\nu}\vec{\rho}_{\mu}\right]\cdot\vec{T}N+\mathrm{H.c.} \, ,
\end{gather}
\begin{gather}
	\mathcal{L}^{1/2+}_{\pi\Delta N^*}=-\frac{3g_{\pi\Delta N^*}}{2f_\pi}\bar{\Delta}^\mu\vec{T}\cdot\partial_\mu\vec{\pi}N^*+{\rm H.c.} \, ,
\end{gather}
\begin{gather}
	\mathcal{L}^{1/2+}_{\gamma N^*\Delta}=-ie\hat{g}_{\gamma\Delta N^*}\bar{\Delta}^\mu\Gamma^{\rm em,\Delta N}_{\mu\nu}T^3N^*A^\nu+{\rm H.c.} \, ,
\end{gather}
where $\Gamma^{\rm em,\Delta}_{\mu\nu}$ is defined in Eq.~(B86) in Ref.~\cite{Kamano:2013iva}
\begin{align}
	\Gamma^{\rm em,\Delta}_{\mu\nu}=&\frac{m_\Delta+m_N}{2m_N}\frac{1}{(m_\Delta+m_N)^2-Q^2}\left[\left(G_M^{\Delta N}-G_E^{\Delta N}\right)3\epsilon_{\mu\nu\alpha\beta}P^\alpha Q^\beta\right.\notag\\
	&+G_{E}^{\Delta N}i\gamma_{5}\frac{12}{(m_{\Delta}-m_{N})^{2}-Q^{2}}\epsilon_{\mu\lambda\alpha\beta}P^{\alpha}Q^{\beta}\epsilon^{\lambda}{}_{\nu\gamma\delta}p_{\Delta}^{\gamma}Q^{\delta}\notag\\
	&\left.+G_C^{\Delta N}i\gamma_5\frac{6}{(m_\Delta-m_N)^2-Q^2}Q_\mu(Q^2P_\nu-Q\cdot PQ_\nu)
	\right] \, ,
\end{align}
with $P=(p_\Delta+p_N)/2$ and $Q=p_\Delta-p_N$.

$\hat{\mathcal{M}}_{\pi N,\gamma N}$: The amplitudes not involving the resonance are the same as in the odd-parity case Eqs.~(\ref{pin1})-(\ref{pin5}), except for the form factor $u^{\rm em}_{+}(k)$ we added. The amplitudes involving the resonances are as follows
\begin{align}
	\mathcal{\hat{M}}^{1/2+}_{s+u,N^*}={}&e \frac{g_{\pi NN^*}}{f_\pi} \frac{\sqrt{3}}{4m_N}
	\left\{\hat{g}_{N\gamma N^{*}}\slashed{k}\gamma_5 \frac{1}{\slashed{q}+\slashed{p}-m_{N^*}^0} (\slashed{q}\slashed{\varepsilon}_\gamma- \slashed{\varepsilon}_\gamma\slashed{q})
	\right.\notag\\ 
	&+(\frac{-\tau_3+3}{2}\frac{g_{p\gamma N^{*+}}}{3}+\frac{\tau_3+3}{2}\frac{g_{n\gamma N^{*0}}}{3})(\slashed{q}\slashed{\varepsilon}_\gamma- \slashed{\varepsilon}_\gamma\slashed{q})\notag\\ &\times\left.\frac{1}{\slashed{p}-\slashed{k}-m_{N^*}^0} \slashed{k}\gamma_5 \right\}u^+(k) \, ,
\end{align}

$\hat{\mathcal{M}}_{\sigma N,\gamma N}$:
\begin{align}
	\mathcal{\hat{M}}_{sN}=-ieg_{\sigma NN}\frac1{\slashed{q}+\slashed{p}-m_N}\Gamma_N u^{\rm em}_{+}(k) \, ,
\end{align}
\begin{align}
	\mathcal{\hat{M}}_{uN}=-ieg_{\sigma NN}\frac1{\slashed{p}-\slashed{k}-m_N}\Gamma_N u^{\rm em}_{+}(k) \, ,
\end{align}

$\hat{\mathcal{M}}_{\pi\Delta,\gamma N}$:
\begin{align}
	\mathcal{\hat{M}}_{sN}=&\sqrt{2}e\frac{f_{\pi N\Delta}}{m_\pi}\varepsilon_\Delta^*\cdot k\frac1{\slashed{q}+\slashed{p}-m_N}\Gamma_N u^{\rm em}_{+}(k) \, ,
\end{align}
\begin{align}
    \mathcal{\hat{M}}_{uN}=&-\frac{2\sqrt{2}}{3}ie\frac{f_{\pi N\Delta}}{m_{\pi}}\tau^3\varepsilon^{\mu *}_{\Delta}\Gamma_{\mu\nu}^{\rm em,\Delta}\varepsilon_{\gamma}^{\nu}\frac1{\slashed{p}-\slashed{k}-m_N}\notag\\
    &\times\slashed{k}\gamma_{5}u^{\rm em}_{+}(k) \, ,
\end{align}
\begin{align}
    \mathcal{\hat{M}}_{u\Delta}=&\frac{\sqrt{2}}{6}(5\tau^3+3)e\frac{f_{\pi N\Delta}}{m_\pi}\varepsilon^{*}_{\Delta\eta}\left[-g^{\eta\mu}\slashed{\varepsilon}_\gamma+\varepsilon^{\eta}_\gamma\gamma^\mu\right]\notag\\
    &\times S^\Delta_{\mu\nu}(p-k)k^\nu u^{\rm em}_{+}(k) \, ,
\end{align}
\begin{align}
    \mathcal{\hat{M}}_{cont}=&\frac{\sqrt{2}}{3}e\frac{f_{\pi N\Delta}}{m_{\pi}}\tau^3\varepsilon_{\gamma}\cdot\varepsilon_{\Delta}^{*} u^{\rm em}_{+}(k) \, ,
\end{align}
\begin{align}
	\mathcal{\hat{M}}_{t\pi}=&-\frac{\sqrt{2}}{3}e\frac{f_{\pi N\Delta}}{m_{\pi}}\tau^3(V_g+Z_g)u^{\rm em}_{+}(k) \, ,
\end{align}
\begin{align}
    \mathcal{\hat{M}}_{t\rho}=&-\frac{f_{\rho N\Delta}}{m_\rho}\frac{g_{\rho\pi\gamma}}{m_\pi}\frac1{\tilde{k}^2-m_\rho^2}\left[\tilde{k}\cdot\varepsilon_\Delta^*\gamma^\mu-\slashed{\tilde{k}}\varepsilon_\Delta^*\right]\notag\\
    &\times\gamma_5\epsilon_{\alpha\beta\eta\mu}q^\alpha\varepsilon_\gamma^\beta\tilde{k}^\eta u^{\rm em}_{+}(k) \, ,
\end{align}
\begin{align}
	\mathcal{\hat{M}}^{1/2+}_{s+u,N^*}=&\left\{-\frac{\sqrt{2}}{4}\frac{eg_{\pi\Delta N^*}\hat{g}_{N\gamma N^{*}}}{f_\pi m_N}\varepsilon_\Delta^*\cdot k\frac1{\slashed{q}+\slashed{p}-m^*_N}(\slashed{q}\slashed{\varepsilon}_\gamma-\slashed{\varepsilon}_\gamma\slashed{q})\right.\notag\\
	&+\frac{\sqrt{2}ie}{3f_\pi}\tau^3(g_{\gamma\Delta^0N^{*0}}+g_{\gamma\Delta^+N^{*+}})g_{\pi NN^*}\varepsilon^{\mu *}_{\Delta}\Gamma_{\mu\nu}^{\mathrm{em,}\Delta}\varepsilon_{\gamma}^{\nu}\notag\\
	&\left.\times\frac1{\slashed{p}-\slashed{k}-m^*_N}\slashed{k}\gamma_{5}\right\}u^+(k) \, ,
\end{align}
where $S_\Delta^{\mu\nu}(p)=\frac1{3(p-m_D)}\left[2(-g^{\mu\nu}+\frac{p^{\mu}p^{\nu}}{m_D^2})+\frac{\gamma^{\mu}\gamma^{\nu}-\gamma^{\nu}\gamma^{\mu}}{2}-\frac{p^{\mu}\gamma^{\nu}-p^{\nu}\gamma^{\mu}}{m_{D}}\right]$, $V_g$ and $Z_g$ are defined in Eqs. (D35) and (D36) in Ref.~\cite{Kamano:2013iva}
\begin{align}
	V_g=&\frac1{2E_{\pi}(k-q)}\frac{\varepsilon_{\Delta}^{*}\cdot k_{1}(k_{1}+k)\cdot\varepsilon_{\gamma}}{E_{N}(q)-E_{\Delta}(k)-E_{\pi}(k-q)}+\varepsilon_{\Delta}^{0*}\varepsilon_{\gamma}^{0} \, ,\\
	Z_g=&\frac1{2E_{\pi}(k-q)}\frac{\varepsilon_{\Delta}^{*}\cdot k_{2}(k_{2}+k)\cdot\varepsilon_{\gamma}}{E-E_{N}(q)-E_{\pi}(k)-E_{\pi}(k-q)+i\epsilon} \, ,
\end{align}
with $k_1=(E_\pi(k-q),\vec{k}-\vec{q})$ and $k_2=(-E_\pi(k-q),\vec{k}-\vec{q})$.
%%%%%%%%%%%%%%%%%%%%%%%%%%%%%%%%%%%%%%%%%%%%%%%%%%

%\bibliographystyle{apsrev4-1}
\bibliography{ref}
\end{document}